\documentclass[final,3p,times,twocolumn]{elsarticle}
 
\usepackage{amssymb}
\usepackage{amsthm}
\usepackage{amsfonts,amsmath}
\usepackage{graphicx,subeqnarray}
\usepackage{color}
\usepackage{changepage}
\usepackage{epsfig}
\usepackage{hyperref}

\journal{Journal of Non-Newtonian Fluid Mechanics}

\begin{document}

\begin{frontmatter}

\title{Microswimming in viscoelastic fluids\\ \small Draft \today }



\author[1,2]{Gaojin Li\fnref{e1}}
\author[3]{Eric Lauga\fnref{e2}}
\author[4]{Arezoo M. Ardekani\fnref{e3}}
\address[1]{State Key Laboratory of Ocean Engineering, Shanghai 200240, China}
\address[2]{School of Naval Architecture, Ocean and Civil Engineering, Shanghai Jiao Tong University, Shanghai 200240, China}
\address[3]{Department of Applied Mathematics and Theoretical Physics, University of Cambridge, Wilberforce Road, Cambridge CB3 0WA UK}
\address[4]{School of Mechanical Engineering, Purdue University, West Lafayette, IN 47907, USA}
\fntext[e1]{Corresponding author: gaojinli@sjtu.edu.cn}
\fntext[e2]{Corresponding author: e.lauga@damtp.cam.ac.uk}
\fntext[e3]{Corresponding author: ardekani@purdue.edu}

\begin{abstract}
The locomotion of microorganisms and spermatozoa in complex viscoelastic fluids is of critical importance in many biological processes such as fertilization, infection, and biofilm formation. Depending on their propulsion mechanisms, microswimmers display various responses to a complex fluid environment:   increasing or decreasing their swimming speed and efficiency, modifying their propulsion kinematics and swimming gaits, and experiencing different hydrodynamic interactions with their surroundings. In this article, we review the fundamental physics of locomotion of biological and synthetic microswimmers in complex viscoelastic fluids. Starting from a continuum  framework, we describe the main theoretical approaches developed to model microswimming in viscoelastic fluids, which typically rely on asymptotically-small dimensionless parameters. We then summarise  recent progress on the mobility of single cells propelled by cilia, waving flagella and rotating helical flagella in unbounded  viscoelastic fluids. We next \textcolor{black}{briefly discuss} the impact of other physical factors, including the micro-scale heterogeneity of complex biological fluids, the role of Brownian fluctuations of the microswimmers, the effect of polymer entanglement and the influence of  shear-thinning viscosity. In particular, for   solution of long polymer chains whose sizes are comparable to the radius of  flagella, continuum models cannot be used and instead Brownian Dynamics    for the  polymers can  predict the swimming dynamics. Finally, we discuss the effect of viscoelasticity on the dynamics of microswimmers  in the presence of surfaces or external flows and its impact 
on collective cellular behavior.
\end{abstract}



\begin{keyword}



\end{keyword}

\end{frontmatter}


\section{Introduction}

\begin{figure*}[t]
\begin{center}
\includegraphics[angle=0,scale=0.3]{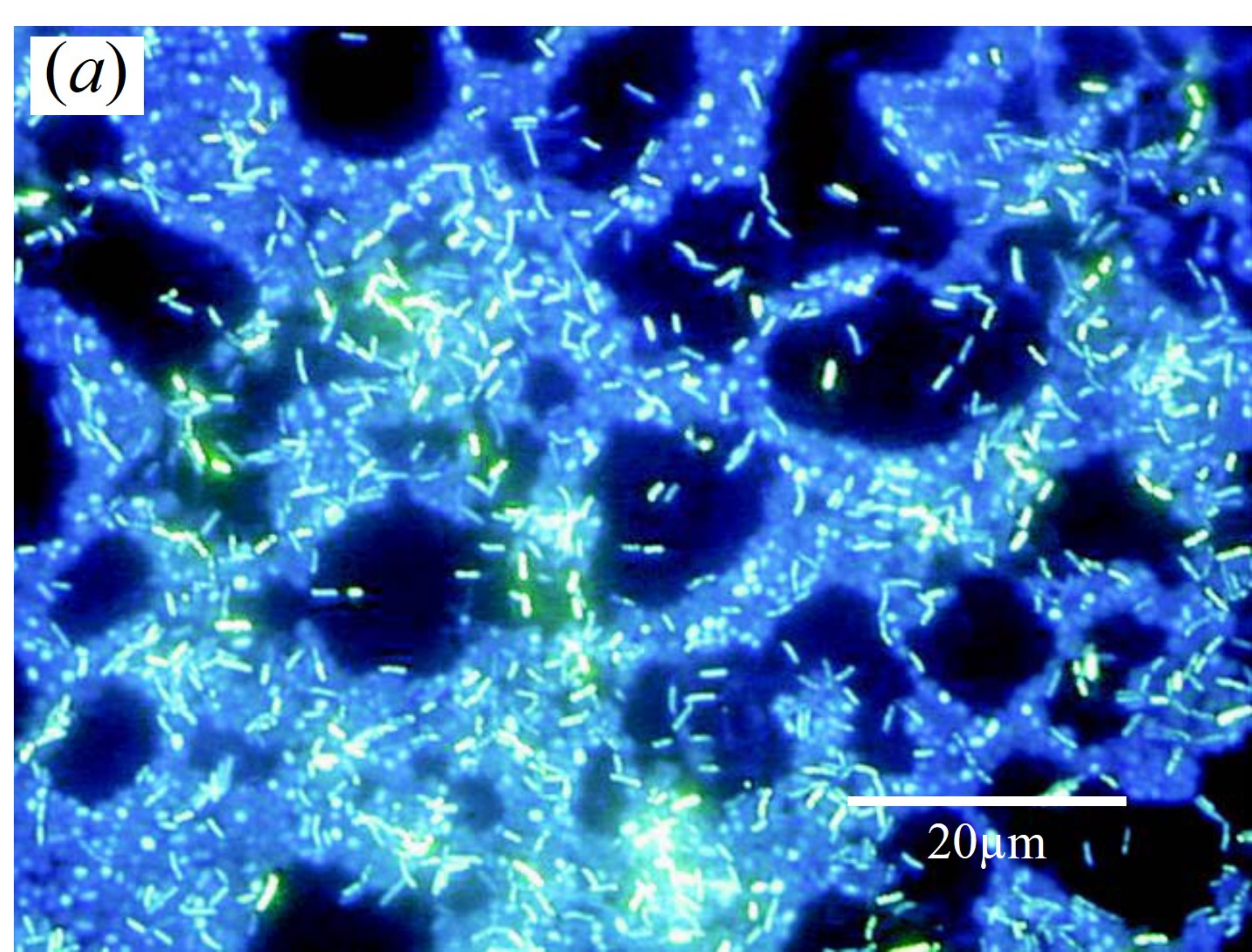}
\includegraphics[angle=0,scale=0.3]{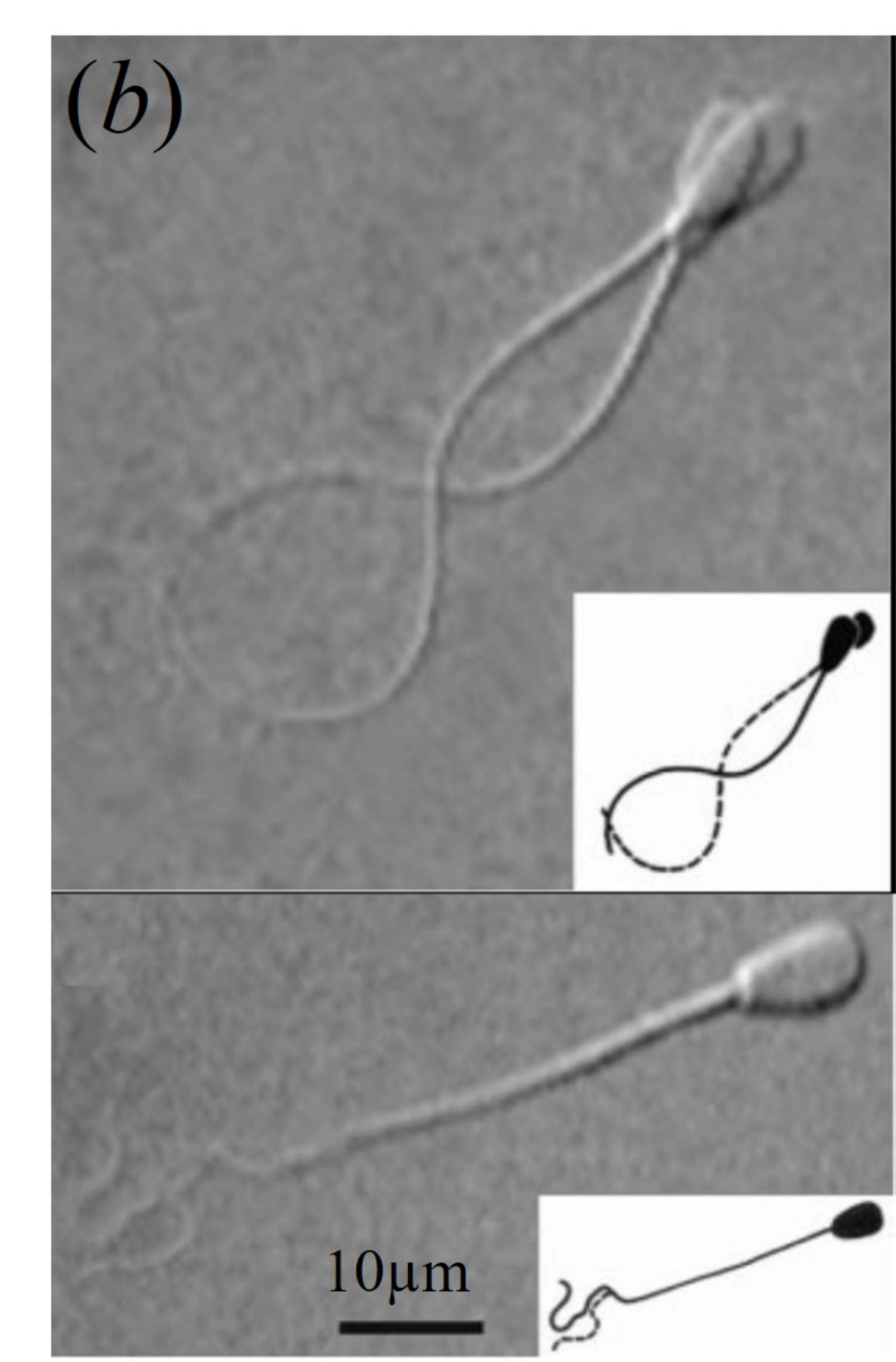}
\includegraphics[angle=0,scale=0.3]{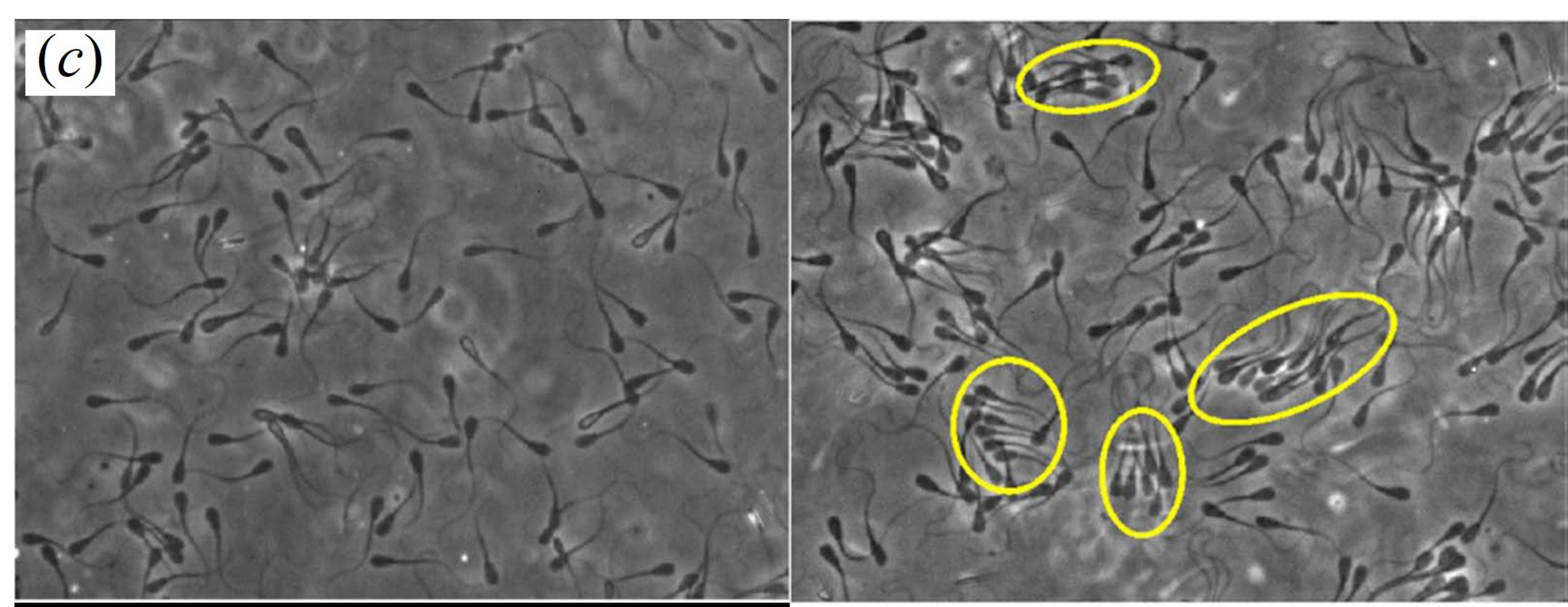}
\caption{Examples of biological microswimming in complex fluids. 
($a$) Polymicrobic biofilm showing heterogenous micro-scale structure  (adapted with permission from Ref.~\cite[]{donlan2002biofilms}). ($b$) Superimposed images of swimming  bull spermatozoa in a Newtonian fluid (top) and a viscoelastic solution of long-chain polyacrylamide (bottom) (adapted with permission from Ref.~\cite[]{ho2001hyperactivation}). ($c$) Bovine spermatozoa  swim in a disordered fashion in a Newtonian fluid (left) but aggregate and swim parallel to each other, while forming clusters in a viscoelastic fluid (right) (adapted with permission from Ref.~\cite[]{tung2017fluid}).}\label{fig:microswimmers}
\end{center}
\end{figure*}

In their natural biological environment, microorganisms and spermatozoa often swim in complex fluids with non-Newtonian characteristics. Examples include mammalian spermatozoa swimming in   viscoelastic mucus in the cervix or along the   fallopian tubes~\cite[]{katz1978movement, katz1980flagellar, katz1981movement, suarez1992hyperactivation, suarez2006sperm, fauci2006biofluidmechanics, hyakutake2015effect}, the bacterium \textit{Helicobacter pylori} moving through the mucus layer covering the stomach and causing ulcers~\cite[]{montecucco2001living, celli2009helicobacter}, the Lyme disease spirochete \textit{Borrelia burgdorferi} penetrating the connective tissues in our skin~\cite[]{moriarty2008real, kimsey1990motility}, and bacteria producing extracellular polymeric substances (EPS) and forming biofilms (figure~\ref{fig:microswimmers}$a$)~\cite[]{o2000biofilm, donlan2002biofilms, costerton1987bacterial, costerton1995microbial, wilking2011biofilms, yazdi2012bacterial, bar2012revised, karimi2015interplay}.
In marine environments, transparent exopolymer particles made of dissolved organic matter or polymer chains released by phytoplankton can initiate bacteria attachment to surfaces and serve as nutrient hot spots~\cite[]{verdugo2004oceanic}. \textcolor{black}{Another important example of a biological flow of complex fluids at the micron scale is the motion of cilia  transporting   layers of mucus  that cover the epithelium of organs and glands. This motion plays a crucial role in the removal of  foreign substances in the mammalian respiratory tracts, the transport of fallopian tube isthmic mucus and ovum, and the transcervical spermatozoa migration in reproductive tracts~\cite[]{fauci2006biofluidmechanics,brennen1977fluid, sleigh1988propulsion, shack1972cilia}.} In addition, ependymal cilia found in specialized brain cells are involved in the transport of cerebrospinal fluid on small scales~\cite[]{o2012analysis}.

In fluid mechanics, the Reynolds number $Re$, which is ratio of inertial force to viscous force, is one of the most important dimensionless parameter. In the context of micro-scale  locomotion at the heart of this paper, we define  $Re\equiv \rho U_c l_c/\mu$, where $l_c$ and $U_c$ are the characteristic length and swimming speed of the microorganism, while $\rho$ and $\mu$ are the mass density and dynamic viscosity of the fluid, respectively.  
Since cells move in a low-Reynolds number world,  $Re \ll1$, when they swim in a Newtonian fluid they must adopt non-reversible strokes to escape the constraints of Purcell's scallop theorem  and achieve self-propulsion~\cite[]{purcell1977life}.

A variety of swimming strategies are employed by motile cells in nature~\cite[]{braybook}. Some protozoa, such as \textit{Paramecium}, and algae, such as \textit{Volvox}, rely on metachronal waves generated by the collective beating motion of their cilia. Many   bacteria, such as the model organism \textit{Escherichia coli} (\textit{E.~coli}), swim by rotating helical flagellar filaments attached to the cell body, while spirochete bacteria, such as \textit{Spiroplasma}, swim by the propagation of rotational waves or shape kinks along their spiral cell bodies. In contrast, most spermatozoa and some nematodes, such as \textit{Caenorhabditis elegans} (\textit{C.~elegans}), swim using a planar waving deformation of their flagella or slender body. In the lab, different types of artificial swimmers have been designed and fabricated to mimic the swimming behavior of microorganisms; this includes
 self-phoretic colloids~\cite[]{moran2017phoretic}, particles externally actuated by magnetic, acoustic, electric fields~\cite[]{dreyfus2005microscopic, wang2014acoustic, mano2017optimal}, vibrated granular matter~\cite[]{deseigne2010collective} and magnetic torque-driven helical robots~\cite[]{peyer2013bio, puente2019viscoelastic}.

To understand the fundamental hydrodynamics at play in swimming cells, a number of theoretical and computational   models have been developed historically. The most widely studied ones include the squirmer model~\cite[]{lighthill1952squirming, blake1971spherical}, Taylor's infinitely-long planar waving sheet~\cite[]{taylor1951analysis} and  waving filament~\cite[]{taylor1952action}, and the rotating helix model~\cite[]{chwang71,higdon1979hydrodynamics}. These simplified models usually prescribe the actuation kinematics of the microswimmers (so-called swimming gait) by assuming that they are not influenced by the surrounding fluid environment. Understanding the impact of complex fluids in these fundamental models has attracted significant research effort in the past decade, much of which will be the focus of this review. For a discussion of the fundamentals of microswimming and  biophysical modeling in a Newtonian fluid, we refer the readers to  review articles~\cite[]{brennen1977fluid, lauga2009hydrodynamics, marchetti2013hydrodynamics, karimi2015interplay, ArdekaniDesai2017modeling, elgeti2015physics, lauga2016bacterial, bechinger2016active, shaebani2020computational} and books~\cite[]{lighthill75,childress81,laugabook}.


The non-Newtonian rheology of complex fluids breaks in general the time-reversal symmetry of the Newtonian Stokes equations. This in turn enables   self-propulsion under  reciprocal motion, as experimentally demonstrated using artificial swimmers~\cite[]{keim2012fluid, qiu2014swimming}. More importantly, complex fluids can significantly impact  the swimming kinematics and motility of individual microswimmers and their collective behavior. The bacterium \textit{E.~coli} swims faster in a polymer solution than   in a Newtonian environment and it follows straighter trajectories with suppressed  wobbling of the cell body~\cite[]{patteson2015running}. The visualization of fluorescently-labeled polymers reveals that the rotation of the flagellar filaments of \textit{E.~coli} strongly stretches the polymers in the fluid and generate local elastic stresses that suppress the unbundling of flagellar filaments, thereby impacting its run-and-tumble dynamics~\cite[]{patteson2015running}. In the case of the algal cell \textit{C.~reinhardtii}, the viscoelasticity of the fluid  changes the wave envelope of the beating flagellar strokes by restricting the displacement of the flagella close to the cell body. The swimming speed of the cell  is then hindered despite the flagella beating at a higher frequency in a polymer solution~\cite[]{qin2015flagellar}. Similar effects are also observed for swimming spermatozoa, whose flagella beat at a higher frequency and with a  smaller amplitude near the cell body in a viscoelastic cervical mucus compared to a Newtonian fluid (see figure~\ref{fig:microswimmers}$b$)~\cite[]{katz1978movement, ishijima1986flagellar, ho2001hyperactivation, smith2009bend}; surprisingly, the  swimming speed of the cells is less affected~\cite[]{katz1978movement}. Collectively, bovine spermatozoa show disorganised individual swimming in Newtonian fluids, but they form dynamic clusters in viscoelastic fluids (figure~\ref{fig:microswimmers}$c$) with cluster sizes and cell-cell alignments increasing with the viscoelasticity of the fluid~\cite[]{tung2017fluid}. These examples demonstrate that the impact of complex fluids on microswimming depends strongly on the type of swimming cells as well as on the interplay between the swimming gait and the rheology of the  fluid environment.

Many non-Newtonian biological fluids, including biofilm and mucus, are hydrated polymeric gels made of $90\%$ water by weight; the remaining components, which are typically conglomeration of extracellular polysaccharides, proteins and lipids, give rise to the elasticity of the fluids and to its shear-thinning viscosity. 
The rheological properties of biofilms show high variability; their elastic modulus $G$ is in the range $10-10^3$~Pa for most biofilms and can reach $10^5$~Pa for biofilms from natural hot springs~\cite[]{stoodley2002biofilm, towler2003viscoelastic, shaw2004commonality, vinogradov2004rheology, rupp2005viscoelasticity}; the  viscosity of biofilms is typically in the range  $\mu\sim10^2-10^5$~Pa$\cdot$s, while their relaxation time is $\lambda\sim10$~s for \textit{S.~epidermidis} biofilm~\cite[]{ hohne2009flexible} and $\sim10^3$~s for a wide range of environmental biofilms~\cite[]{shaw2004commonality}. In the case of mucus, we have $\mu\sim10^{-2}-10^2$~Pa$\cdot$s, $G\sim0.1-10^2$~Pa, and $\lambda\sim1-10$~s~\cite[]{litt1976mucus, lai2009micro}. Since the relevant  time scale for microswimming \textcolor{black}{along the size of the swimmer itself} is $l_c/U_c\sim0.1-1$~s and the flagellum beating period is $T\sim10^{-2}-0.1$~s, we see that the relaxation time of these complex fluids is comparable to, or much larger than, the time scales of microswimming. It is thus expected that  microswimmers should be strongly influenced by  elastic stresses in a complex fluid and thus be subject to the full range of non-Newtonian effects, including stress relaxation, strain retardation, and  normal stress differences. A review of these classical topics can be found in  textbooks on non-Newtonian fluids, including the different constitutive equations models that have been used by the community to model swimming in polymeric fluids (second-order fluids, Oldroyd-B, Giesekus, FENE models)~\cite[]{bird1987dynamics, morrison2001understanding}.

While many studies model complex fluids as  continuum media,  it is important to realize that most biological examples   are highly heterogeneous at the microscopic scale~\cite[]{stewart2008physiological}. Many microorganisms experience    a biological environment of entangled fibers of around $10-100$~nm in diameter and pores ranging in size from 0.1 to $1$~$\mu$m~\cite[]{lai2010nanoparticles}. For example, an aerobic biofilm is a porous structure with polymer conglomerations connected by voids and channels of about $10-100~\mu\textrm{m}$ in diameter that enhance the transport of oxygen and nutrients through the colony~\cite[]{de1994effects}. Such micro-scale heterogeneity  impacts the motility of microswimmers. In dermis and gelatin matrices, the Lyme disease spirochete \textit{B.~burgdorferi} may temporally adhere to the matrix and undergo wriggling and lunging movements that are not observed in a liquid medium~\cite[]{harman2012heterogeneous}. In a semi-dilute polymer solution, the bacterium \textit{E.~coli} swims faster than in a Newtonian fluid because its rotating flagella strongly stretches the polymer molecules in its vicinity and thus carves out a low-viscosity tunnel free of polymers around the flagellum. \textcolor{black}{The flagellum motor gains a larger rotation speed and therefore increases the  speeds of the  cells}~\cite[]{martinez2014flagellated, zhang2018reduced, zottl2019enhanced} 
Microorganisms may also explicitly change the local environment to assist their motility in complex biological fluids. For example, in the acidic environment of the stomach, the pathogenic bacterium \textit{H.~pylori}  increases the local pH by secreting an enzyme  that transforms the viscoelastic mucin gel into a viscous liquid thereby facilitating the penetration of the cell through the  epithelium  surface of the stomach~\cite[]{celli2009helicobacter, bansil2013influence}. In cervical mucus, ``following'' spermatozoa  swim slower than   ``vanguard'' ones despite having the same flagellar beating frequencies and shapes. Such a difference in propulsive efficiency is due to the alteration of the local mucus properties resulting from the sustained permeation by spermatozoa~\cite[]{katz1981movement}. To better understand the impact of micro-scale heterogeneity,   recent studies have thus proposed to use    coarse-grained polymer models~\cite[]{zhang2018reduced, zottl2019enhanced}.

Beyond microorganism swimming, the  motion of small-scale bodies  in complex fluids is encountered in   other related settings. For example, in microrheology, the local mechanical properties of a soft, complex material can be extracted from the thermal or forced motion of colloidal probes~\cite[]{waigh2005microrheology, squires2010fluid, khair2010active}. Another example is that of the electrophoresis of charged colloidal particles or macromolecules in gel matrices, which is a common biochemical technique to separate biomolecules   based on their size and charge~\cite[]{hames1998gel}, and for which some of the models developed to address microswimming are applicable. 

In this article, we review recent theoretical, computational and experimental progress on  biological and synthetic microswimming in complex fluids. Our work follows naturally from previous reviews devoted to active colloids in complex fluids~\cite[]{patteson2016active} and to the role of complex fluids in biological systems~\cite[]{spagnolie2015complex, martinez2021active}. \textcolor{black}{Our paper, written   in  celebration of the legacy  of J.~G.~Oldroyd, focuses on  the impact of  fluid viscoelasticity on microswimming, and includes both individual motion as well as hydrodynamic interactions in viscoelastic fluids (typically less covered in previous reviews). We  discuss as well the influences of non-continuum effects and thermal noise, both of which may be  important at the micron scale. Polymer entanglement and the role of shear-thinning viscosity are also briefly mentioned.} 

\textcolor{black}{This review is organized as follows.} We first summarize the key dimensionless parameters and the relevant theoretical continuum framework in Section~\ref{sec:theory}. We next discuss in Section~\ref{sec:indiv}    the impact of viscoelasticity on the motility of  individual microswimmers. We then focus in Section~\ref{sec:int}   on the role of hydrodynamic interactions in complex fluids: swimmer/surface, swimmer/flow and swimmer/swimmer. Finally  we  conclude  in Section~\ref{sec:conclusion} with our perspective on the direction of research in this area.

\section{Theoretical framework}
\label{sec:theory}
\subsection{Parameters and dimensionless numbers}

\begin{figure}[t]
\begin{center}
\includegraphics[angle=0,scale=0.6]{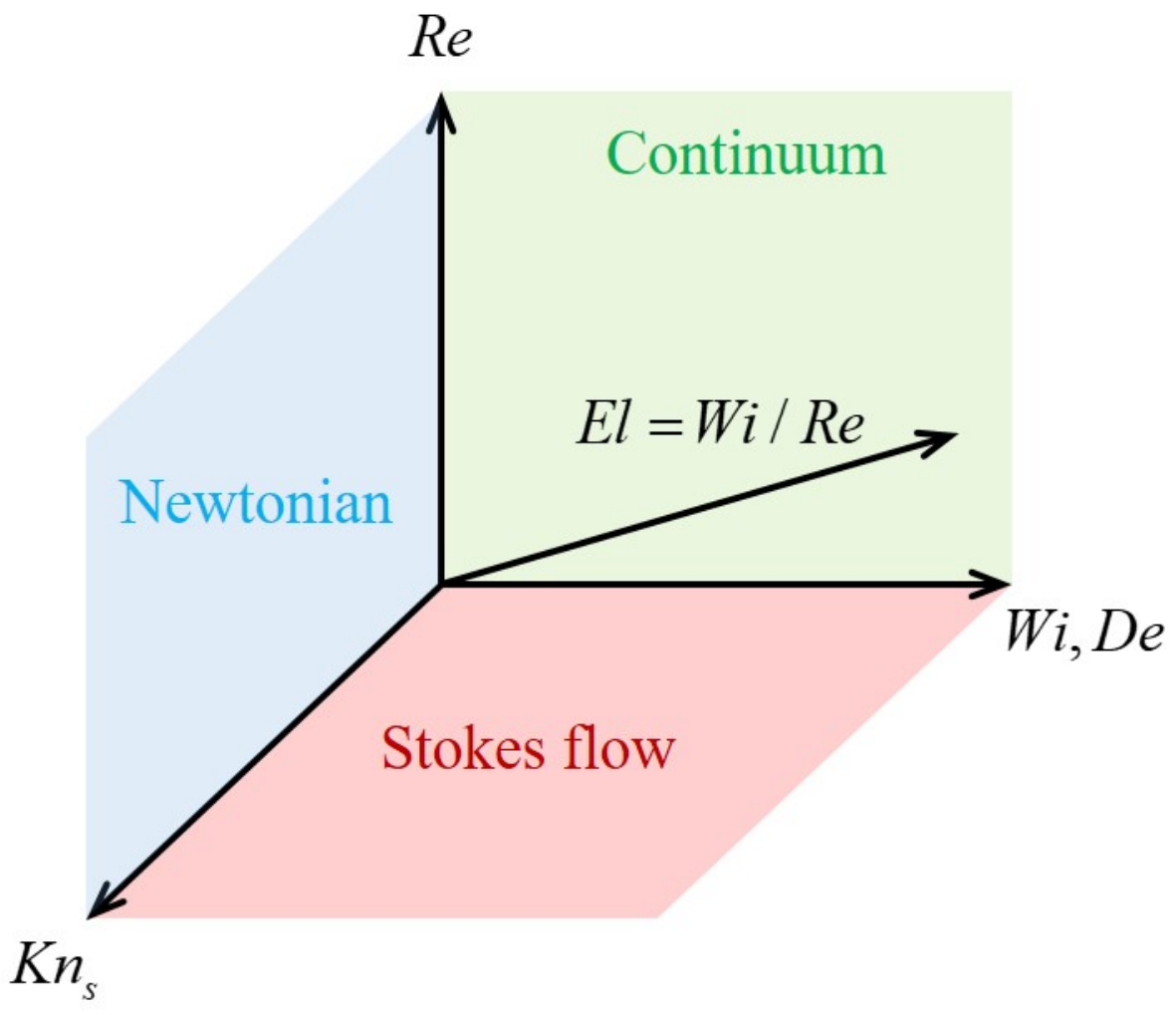}
\caption{The space of dimensionless parameters relevant to microswimming in viscoelastic fluids: Reynolds number $Re$,   Deborah number $De$,    Weissenberg number $Wi$,   elasticity number $El$ and   Knudsen number $Kn_s$.}\label{fig:parameters}
\end{center}
\end{figure}

We start by introducing the classical dimensionless numbers governing fluid dynamics in a viscoelastic (polymeric) fluid and the governing equations relevant for  microswimming. We denote by $\lambda$   the polymer relaxation time, while $t_c, \dot{\gamma}_c$ and $l_c$ are the characteristic time, shear rate and length scales for the flow around the microswimmer. The mass density of the fluid is $\rho$, $\mu=\mu_p+\mu_s$ is the total fluid viscosity, while $\mu_s$ and $\mu_p$ are the viscosity of the solvent and polymer, respectively.

Beyond the Reynolds number $Re$, the key dimensionless parameters for the fluid flow are the Deborah number $De$, Weissenberg number $Wi$, elasticity number $El$, the dimensionless polymer and solvent viscosities $\hat{\mu}_p$ and \textcolor{black}{$\hat{\mu}_s$,  the suspension Knudsen number $Kn_s$, and the translational and rotational P\'eclet number,  $Pe_t$, $ Pe_r$, respectively.} They are classically defined as
\begin{subequations}\label{eq:parameter}
\begin{equation}
De=\lambda/t_c,
\end{equation}
\begin{equation}
Wi=\lambda\dot{\gamma}_c,
\end{equation}
\begin{equation}
El=Wi/Re=\lambda\mu/\rho l_c^2,
\end{equation}
\begin{equation}
\hat{\mu}_p=\mu_p/\mu,\quad\hat{\mu}_s=1-\hat{\mu}_p,
\end{equation}
\begin{equation}
Kn_s=l_d/l_c,
\end{equation}
\begin{equation}\label{eq:Pe}
Pe_t=l_cU_c/D_t,\quad Pe_r=U_c/D_rl_c.
\end{equation}
\end{subequations}

The Deborah number $De$ represents the ratio between the polymer relaxation time and observation time of the fluid flow. For a swimming microorganism, the flow time scale is typically defined as either $t_c=1/\dot{\gamma}_c=l_c/U_c$ or $t_c=1/\omega$, where $U_c$ is the swimming speed and $\omega$ is the frequency for a periodic motion of the swimmer, for example the waving motion of the flagella. 

The Weissenberg number $Wi$ is the \textcolor{black}{product of the polymer relaxation time $\lambda$ with the characteristic flow shear rate $\dot{\gamma}_c$. It also measures} the ratio between the elastic force (first normal stress difference in a steady simple shear flow) $\sim\lambda\mu\dot{\gamma}_c^2$ \cite[]{poole2012deborah} and the viscous force $\sim\mu\dot{\gamma}_c$. In some flows, e.g.~steady squirming motion, we have  $Wi\sim De$, while in general these two dimensionless numbers  can be quite  different. \textcolor{black}{For example, the unsteady viscoelastic flow induced by an oscillatory shear on a squirmer surface can have different magnitudes for $De$ and $Wi$. In the classical oscillatory shear flow of viscoelastic fluids, the Pipkin diagram classifies various regimes as `viscometric' for $De\ll1$, `linear viscoelastic' for $Wi\ll1$, and `nonlinear viscoelastic' for non-small $De$ and $Wi$~\cite[]{giacomin1993large}.}

\textcolor{black}{The Weissenberg and Deborah numbers are often related to one another via a length scale relevant to the microswimmer. For example, in the case of a self-propelled phoretic particle, the shear rate inside the thin interaction layer  is much stronger than that in the flow in the bulk region and we have $Wi=De/\delta$ with $\delta\ll1$ being the dimensionless boundary layer thickness~\cite[]{li2020electrophoresis}.}
In the case of a waving flagellum, the ratio between the Deborah and Weissenberg numbers depends on the dimensionless beating amplitude,  $Wi=A De$, with  $A=A^*/l_c$, where $A^*$ is the dimensional amplitude of the wave. However, even at small-amplitude, the nonlinear impact of viscoelastic stress is important~\cite[]{lauga2007propulsion}.

The third dimensionless number, termed the elasticity number and defined as $El=Wi/Re$, is independent of the swimming kinematics and is only a function of the fluid properties and the \textcolor{black}{size} of the swimmer. It corresponds to the ratio of the polymer relaxation time $\lambda$, to the time scale for momentum diffusion, $\sim l_c^2/(\mu/\rho)$. \textcolor{black}{For typical microswimmers in biofilms and mucus, we have $El\gg1$.} 

The dimensionless polymer viscosity, $\hat{\mu}_p$, characterizes the ratio of the polymeric viscosity to the solution viscosity $\mu$ (which includes both polymer and solvent), while $\hat{\mu}_s$ is the solvent viscosity normalized by the solution viscosity. Consequently, $\hat{\mu}_p+\hat{\mu}_s=1$. For many classical constitutive models, such as Oldroyd-B, Giesekus and the FENE family of models,  $\hat{\mu}_s=\lambda_r/\lambda$ is the ratio between the retardation time and the relaxation time of the polymer. Note that the viscosity in an Oldroyd-B model is constant in a shear flow, whereas it displays a shear-thinning behavior in a Giesekus fluid or FENE-P fluid.

\textcolor{black}{In a dilute polymer solution, i.e.~one for which the polymer concentration $c$ is lower than the overlap concentration, $c<c^*$, the polymer viscosity is the product of the concentration with the intrinsic viscosity of the polymer, $\mu_p\sim c[\eta]$ with $[\eta]\sim1/c^*\sim M^{3\nu-1}$ follow a power-law dependence on the polymer molecular weight $M$; $\nu=3/5$ and $1/2$ are the scaling exponents of polymer solutions of good and athermal solvents, respectively.   
In contrast, for semi-dilute or entangled polymer solutions, the polymer viscosity scales as $\mu_p\sim M^a$, and it  grows quickly with increasing polymer molecular weight since $a=3.4-4$~\cite{rubinstein2003polymer}.
For relevant biological fluids, the dimensionless polymer viscosity $\hat{\mu}_p$ can span a wide range of values, from 0.01 to around 1.}

Next, the dimensionless suspension Knudsen number $Kn_s$ \textcolor{black}{quantifies non-continuum aspects} and measures   the ratio between the size of the dispersed phase $l_d$ and the characteristic swimming flow length scale $l_c$. It is sometimes called the gradient number to reflect the fact that the two ends of a long bead-spring dumbbell experience non-local flow gradient~\cite[]{zhang2018reduced}. In a viscoelastic polymer solution, the relevant value for $l_d$ is the radius of gyration of the polymer, \textcolor{black}{while in a solution of entangled polymers, $l_d$ is  the typical mesh size of the  network.} In a suspension of rigid colloids, $l_d$ is the typical size of the  colloids. When $Kn_s\ll1$, the complex fluid can be modelled as a continuum around the microswimmer, otherwise it is not a continuum medium and the details of its microstructure need to be taken into account.

\textcolor{black}{Finally, the P\'eclect numbers, $Pe_t$ and $Pe_r$, characterize the ratio between the swimming motion and the translational ($Pe_t$) or rotational Brownian motion ($Pe_r$) of the swimmer due to the thermal fluctuations or swimmer's own random reorientation (such as the tumbling motion of flagellated bacteria, e.g.~{\it E.~coli}), where in Eq.~(\ref{eq:Pe}) $D_t$ and $D_r$ are the translational and rotational diffusivities of the swimmer.} These two parameters, which are used in  mean-field theories, play an important role in the  collective motion of microswimmers~\cite[]{pedley2010instability, kasyap2012chemotaxis}. 
 
Depending on the constitutive models, other dimensionless parameters also affect  microswimming in viscoelastic fluids. This includes the mobility factor $\alpha_m$ in the Giesekus model, which measures the anisotropic hydrodynamic drag on the polymer molecules, and the maximum dimensionless length of the polymers in the FENE-type models. 

 \subsection{Governing equations}

The motion of a microswimmer in a non-Newtonian fluid is controlled by the interplay between the low-Re hydrodynamics governed by the Cauchy momentum equation, the swimming kinematics or the internal actuation mechanism of the swimmer (which provides the boundary condition for the fluid), and the fluid rheological characteristics (i.e.~the constitutive equation for the complex fluid). Classical governing equations for continuum viscoelastic fluids can be written as~\cite[]{lauga2009life, lauga2014locomotion}
\begin{subeqnarray}\label{eq:governing}
\nabla p&=&\nabla\cdot\mbox{\boldmath$\tau$},\slabel{eq:governinga}\\
(1+\mathcal{A})\mbox{\boldmath$\tau$}+\mathcal{M}(\mbox{\boldmath$\tau$}, \mbox{\boldmath$u$})
&=&\mu(1+\mathcal{B})\dot{\mbox{\boldmath$\gamma$}}+\mathcal{N}(\dot{\mbox{\boldmath$\gamma$}}, \mbox{\boldmath$u$}),\quad \slabel{eq:governingb}
\end{subeqnarray}
where $p$ is the dynamic pressure, $\mbox{\boldmath$\tau$}$ is the  deviatoric part of the stress tensor, which includes both Newtonian and non-Newtonian contributions to the stress. The Cauchy momentum in the absence of inertia, Eq.~\eqref{eq:governinga}, states therefore simply that the divergence of the total stress tensor, ${\boldsymbol\sigma} = -p{\bf 1} +{\boldsymbol\tau} $, is zero. In Eq.~\eqref{eq:governingb},  $\mu$ is the total zero-shear rate viscosity of the fluid,  $\dot{\mbox{\boldmath$\gamma$}}=\nabla\mbox{\boldmath$u$}+\nabla\mbox{\boldmath$u$}^T$ is the shear rate tensor, while  $\mathcal{A}$ and $\mathcal{B}$ are two linear differential operators in time that capture the polymer relaxation and retardation. In contrast, the symmetric nonlinear operators
$\mathcal{M}$ and $\mathcal{N}$ represent convection, stretching and nonlinear relaxation; they include in particular all terms arising from the objective derivatives of the deviatoric stress and of the shear rate tensor. The constitutive relationship stated in Eq.~(\ref{eq:governingb}) describes all the classical differential constitutive relationship for polymeric fluids, including all Oldroyd models. The equations  above are then to be solved with  kinematic and/or stress boundary conditions provided by the swimmer; in parallel, the dynamics of a microswimmer is subject to instantaneous force-free and torque-free conditions, allowing to determine in a quasi-steady fashion its  linear and angular swimming velocities.

 \subsection{Reciprocal theorem}

In  cases where the swimmer velocity is of   interest and where the swimmer moves by imposing velocity boundary conditions, one can use the Lorentz' reciprocal theorem (a form of the principle of virtual work~\cite[]{stone1996propulsion, elfring2015note}) to determine the velocity in some asymptotic limits. 
Rather than directly solving for the flow field, the reciprocal theorem uses a dual problem for  the swimmer with the same instantaneous shape undergoing  rigid-body motion. Such theoretical analysis using the reciprocal theorem typically requires an asymptotic expansion in a small parameter~\cite[]{lauga2007propulsion, lauga2009life, datt2017active}, although it can be shown to also work  for a strictly tangential squirming motion in a linear viscoelastic fluid~\cite[]{lauga2014locomotion}. In these cases, \textcolor{black}{Eq.~(\ref{eq:governing}$b$) takes a simpler form and} the deviatoric stress can always be written as
\begin{equation}\label{eq:smalleps}
\mbox{\boldmath$\tau$}=\mu\dot{\mbox{\boldmath$\gamma$}}+\epsilon\mbox{\boldmath$\Sigma$}(\mbox{\boldmath$u$}),
\end{equation}
where $\epsilon$ is a small dimensionless parameter and the symmetric tensor $\mbox{\boldmath$\Sigma$}$ is an explicit nonlinear function of the velocity field $\mbox{\boldmath$u$}$. 
The small parameter $\epsilon$ in Eq.~\eqref{eq:smalleps} could be the Deborah number $De$ in a weakly nonlinear polymeric fluid, the viscosity ratio $\hat{\mu}_p$ in a dilute polymer solution, or the small amplitude $A$ of the swimming kinematics.  For a generalized linear viscoelastic fluid, the above constitutive relationship is expressed in Fourier domain and $\mu$ is the frequency-dependent complex viscosity~\cite[]{lauga2014locomotion}. \textcolor{black}{Note that the results obtained with the generalized linear viscoelastic model should be used with caution since that model is not objective (i.e.~frame invariant)~\cite{bird1987dynamics}.}

One can then apply the reciprocal theorem at each order in $\epsilon$ in order to determine the velocity of the swimmer asymptotically. We assume that the swimmer imposes a known swimming gait velocity $\mbox{\boldmath$u$}_{s1}$ on its  surface $S$, which is taken to be either the fixed reference surface   for the small-amplitude perturbations ($\mbox{\boldmath$u$}_{s1}$ is the leading order contribution) and the instantaneous body surface for other perturbations ($\mbox{\boldmath$u$}_{s1}=\mbox{\boldmath$u$}_{s}$ is the imposed velocity).  At the leading order (O(1) for $De,\hat{\mu}_p\ll1$ and O($\epsilon$) for $A\ll1$), the reciprocal theorem then leads to
\begin{equation}\label{eq:reciprocal1}
\hat{\mbox{\boldmath$F$}}\cdot\mbox{\boldmath$U$}_1
+\hat{\mbox{\boldmath$L$}}\cdot\mbox{\boldmath$\Omega$}_1
=-\iint_{S}\mbox{\boldmath$n$}\cdot\hat{\mbox{\boldmath$\sigma$}}\cdot\mbox{\boldmath$u$}_{s1}dS,
\end{equation}
where the hat refers to the variables for the dual problem, $\hat{\mbox{\boldmath$F$}}$ and $\hat{\mbox{\boldmath$\Omega$}}$ are the hydrodynamic force and torque exerted on the swimmer  undergoing   instantaneous rigid-body motion, while $\mbox{\boldmath$U$}_1$ and $\mbox{\boldmath$\Omega$}_1$ are the leading-order linear and angular velocities  of the swimmer. \textcolor{black}{Note that the reciprocal theorem can also be written using resistance tensors~\cite[]{elfring2015note,elfring2017force}. In practice, dual problems with force- and torque-free conditions are often considered separately to determine the swimmer translational and rotational speeds.}

The expression in Eq.~\eqref{eq:reciprocal1} is   the same as the one obtained for a Newtonian fluid~\cite[]{stone1996propulsion}, which leads to     three conclusions: (1) In a generalized linear viscoelastic fluid, a swimmer with a fixed body shape undergoing a tangential squirming motion has the exact same swimming velocity and rotation rate as in a Newtonian fluid; (2) In a weakly non-Newtonian fluid, the  velocity of the swimmer at O(1)  is the same as the one in a Newtonian fluid, so possible changes to the swimming kinematics can only occur at O($De$) or O($\hat{\mu}_p$) (or later); (3) For small-amplitude \textcolor{black}{reciprocal} swimming, the time-averaged locomotion at this order is always zero $\langle\mbox{\boldmath$U$}_1\rangle=\langle\mbox{\boldmath$\Omega$}_1\rangle=0$ because the time average of the  swimming gait $\langle\mbox{\boldmath$u$}_{s1}\rangle\equiv0$ on the fixed reference surface $S$.

At next order (i.e.~O($\epsilon$) for $De,\hat{\mu}_p\ll1$ and O($\epsilon^2$) for $A\ll1$), the linear and angular velocities of the swimmer $\mbox{\boldmath$U$}_2$ and $\mbox{\boldmath$\Omega$}_2$ satisfy
\begin{equation}\label{eq:reciprocal2}
\hat{\mbox{\boldmath$F$}}\cdot\mbox{\boldmath$U$}_2
+\hat{\mbox{\boldmath$L$}}\cdot\mbox{\boldmath$\Omega$}_2
=-\iint_{S}\mbox{\boldmath$n$}\cdot\hat{\mbox{\boldmath$\sigma$}}\cdot\mbox{\boldmath$u$}_{s2}dS
+\iiint_{V}\mbox{\boldmath$\Sigma$}(\mbox{\boldmath$u$}_1):\nabla\hat{\mbox{\boldmath$u$}}dV,
\end{equation}
where $V$ is the fluid domain instantaneously surrounding the body of the swimmer and $\mbox{\boldmath$u$}_{s2}$ is the higher-order swimming gait  (this term only contributes in the case of a  small-amplitude expansion).

The result in Eq.~(\ref{eq:reciprocal2}) states that the  swimming kinematics at next order ($\mbox{\boldmath$U$}_2$ and $\mbox{\boldmath$\Omega$}_2$)  are induced by two contributions from the right-hand side of the equation: (i) the imposed  velocity on the swimmer surface at the next-order, which is of the  same form as the Newtonian contribution from Eq.~(\ref{eq:reciprocal1}), and (ii) a new non-Newtonian term given by the virtual work of the nonlinear stress   against the strain-rate tensor in the dual problem.

In this new term, computing the tensor $\mbox{\boldmath$\Sigma$}$ from the 
leading-order flow field $\mbox{\boldmath$u$}_1$ is the key step. In  many cases, this can only been done numerically. For  small-amplitude periodic swimming, that term  can be first evaluated in Fourier domain and then inverse Fourier-transformed back to the physical domain.  For a swimmer deforming with small amplitude $A\ll1$, the surface and volume integrals in Eq.~(\ref{eq:reciprocal2}) are then expanded for an undeformed body shape. For $A\ll1$, since the leading-order time-averaged velocity at O($A$) is zero, the swimming motion due to the non-Newtonian effect is of  order O($A^2$), which is the same  as the Newtonian contribution. In contrast,  for a swimmer in a weakly non-Newtonian fluid,   non-Newtonian stresses impact the velocity at order O($De$) or O($\hat{\mu}_p$).

 \begin{figure*}[t!]
\begin{center}
\includegraphics[angle=0,scale=0.7]{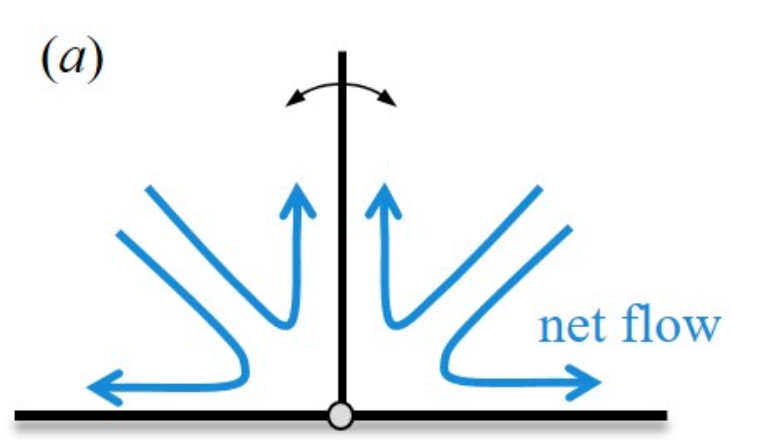}
\includegraphics[angle=0,scale=0.7]{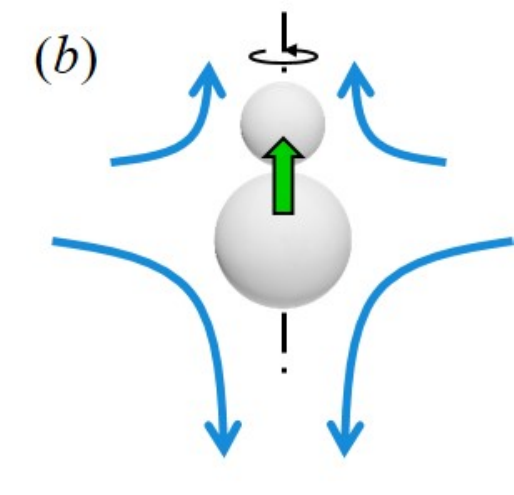}
\includegraphics[angle=0,scale=0.7]{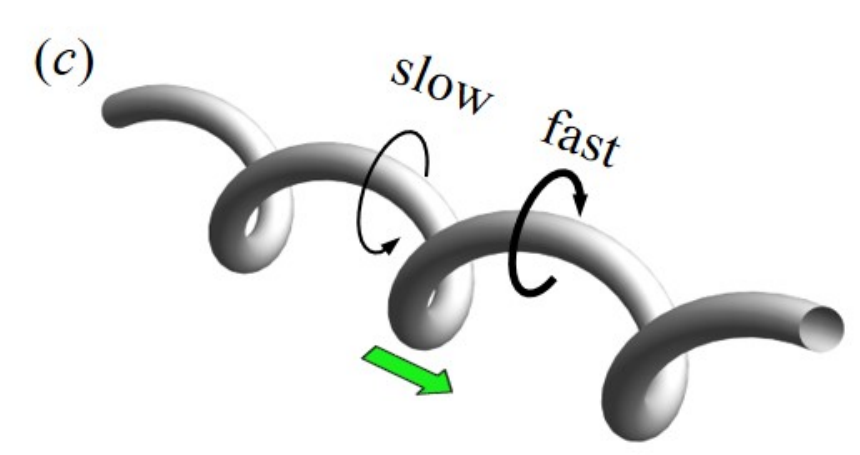}
\caption{Examples of reciprocal swimmers in viscoelastic fluids.  ($a$) A periodically flapping plate tethered to a wall generates net pumping flows  in a non-Newtonian fluid ~\cite[]{pak2010pumping}. ($b$) An asymmetric snowman swimmer rotating around its axis is subject to net normal stresses differences  in a viscoelastic fluid and as a result swims in the direction from the large to the small sphere ~\cite[]{pak2012micropropulsion}. ($c$) A helical filament undergoing periodic backward-and-forward traveling waves at different speed moves in a viscoelastic fluid in the direction of the fast wave ~\cite[]{fu2009swimming}.}
\label{fig:reciprocal}
\end{center}
\end{figure*}

While Eq.~(\ref{eq:reciprocal2}) can be used to calculate the velocity of a free swimmer in a viscoelastic fluid  asymptotically,  in the case of large-amplitude swimming and in a strongly non-Newtonian fluid, direct numerical integration of Eq.~(\ref{eq:governing}) is required to capture the full nonlinear effect of the fluid viscoelasticity. 
In the next section, we overview  recent  progress in our understanding of the motion of individual swimmer models in viscoelastic fluids.

\section{Motility of individual swimmers in viscoelastic fluids}
\label{sec:indiv}

In this section, we discuss the dynamics of a single swimmer in an unbounded viscoelastic fluid. The two most fundamental questions are: (i) for Newtonian non-swimmers, can viscoelasticity lead to locomotion? and (ii) for Newtonian swimmers, does the viscoelasticity of the fluid  enhance or impede microswimming? As we see below, while the answer to (i) is broadly yes due to the generation of nonlinear stresses, the answer to (ii) is not obvious but depends on various factors, including the geometry of the swimmer, its swimming gait, the dimensionless parameters of the flow, and the values of $De$ and $Wi$. 

\subsection{Reciprocal swimmer}
In a Newtonian fluid, microswimming is constrained by Purcell's scallop theorem~\cite[]{purcell1977life}, stating that reciprocal deformation of a deforming body (i.e~one for which the sequence of shapes is identical under a time-reversal symmetry) leads to  zero net propulsion on average due to the linearity and time reversibility of the Stokes equations. Indeed, as seen in Eq.~(\ref{eq:reciprocal1}), time only enters the problem as a parameter in the boundary conditions so that the locomotion of any swimmer in a Newtonian fluid is instantaneously and linearly determined by the velocity boundary condition $\mbox{\boldmath$u$}_{s1}$. As a consequence,  the total distance travelled by a swimmer over a period is independent of the 
rate of actuation of the swimmer body but depends only on the  sequence of body shapes. 
In contrast, complex fluids display nonlinear rheological properties (including shear-dependent viscosities and normal-stress differences) and therefore the stresses exerted on any swimmer are expected to break the constraints of the scallop theorem and to enable net propulsion in general.

Two types of reciprocal swimming in complex fluids have been demonstrated. The first  type consists of swimming enabled by normal-stress differences arising in shear flows.  
For a tethered semi-infinite rigid plate undergoing   small-amplitude sinusoidal motion with flapping amplitude $A$  (figure~\ref{fig:reciprocal}$a$), an asymptotic analysis showed that \textcolor{black}{the non-Newtonian stresses generate a net flow under time-periodic actuation, and the flapping motion} leads to non-Newtonian propulsive forces exerted  on the plate at O($A^2$)~\cite[]{normand2008flapping}. In contrast,  a net pumping flow is induced at O($A^4$)~\cite[]{pak2010pumping}, with a strength that has a non-monotonic dependence on the Deborah number~\cite[]{pak2010pumping}.

Experimentally, the viscoelastic propulsion of reciprocal flapping swimmers  was demonstrated using a magnetically polarized 
dumbbell-swimmer in an oscillating magnetic field, with net motion occurring for swimmers with asymmetric shapes or for symmetric swimmers near a wall in viscoelastic fluids, with no motion in a Newtonian fluid~\cite[]{keim2012fluid}.

A related example of a swimmer driven by normal stress differences in viscoelastic fluids is the dumbbell (`snowman') swimmer rotating around its symmetry axis~\cite[]{pak2012micropropulsion, puente2019viscoelastic}.  The rotation of the spheres in the dumbbell induces elastic hoop stresses along the curved streamlines in the azimuthal  direction, leading to an inward contraction in the radial direction and thus secondary \textcolor{black}{streaming} flows which push the two spheres   away from each other. The strengths of these flows are unbalanced if the two spheres have different sizes and hence for an asymmetric dumbbell they lead to a net motion (towards the side of the small sphere, see figure~\ref{fig:reciprocal}$b$). A similar mechanism  leads to the propulsion of an axisymmetric swimmer performing a harmonic torsional oscillation in a viscoelastic fluid~\cite[]{bohme2015propulsion}. \textcolor{black}{Due to elasticity-induced streaming flows, a reciprocal swimmer composed of two spheres of unequal sizes which oscillate periodically along its axis of symmetry can self-propel in a viscoelastic fluid \cite[]{datt2018two}. For both in-phase and anti-phase oscillations, the net motion is always in the direction of the smaller sphere. A swimmer composed of two  spheres that can deform elastically is seen to swim in the same direction in a Newtonian fluid \cite[]{datt2018two}. Recently, the motion of a torque-free swimmer composed of two spheres rotating around their symmetry axis in opposite directions was studied~\cite[]{binagia2021self}. The swimmer is seen to always translate towards the larger sphere (the one with the slower rotation), presumably because the fast rotation of the smaller sphere creates a stronger streaming flow directed away from the larger sphere. 
Other shapes of revolution and  arrangements of spheres were also studied, leading to translation possible in both directions.
 }

A second type of reciprocal microswimming in viscoelastic fluids does not exploit steady-state normal stresses but instead relies on the rate-dependance of the flows and forces generated in the fluid. Indeed, unlike the Newtonian limit, the instantaneous swimming speed in a viscoelastic fluid is expected to depend nonlinearly on the rate at which the swimmers deforms its body, and therefore in general an asymmetry in a periodic actuation should result in an asymmetry in swimming kinematics.  This strategy was proposed in Ref.~\cite[]{fu2009swimming}  using  the example of a helically waving filament propagating waves at different speeds in  forward and backward strokes (specifically the wave moved  forward by one wavelength in the first third of the time period and then backward by one wavelength in the final two thirds of the period, see figure~\ref{fig:reciprocal}$c$).   In both periods, the helical filament moved in the direction opposite to the  wave propagation but with a smaller total displacement  in the first (forward) cycle  due to a larger value of $De$. Since no locomotion occurs in the Newtonian limit $De=0$ and propulsion is expected to be poor in the high-$De$ limit, net swimming is maximal for $De\sim\textrm{O}(1)$~\cite[]{fu2009swimming}. \textcolor{black}{Note that in this case, the velocity of the flagellum  is not simply obtained as the sum of the velocities of each stroke because the time-history of the polymer plays an important role.}

\subsection{Squirmer model}

A popular model in the theoretical treatment of swimming at low Reynolds number is the  squirmer model~\cite[]{lighthill1952squirming,blake1971spherical}. A squirmer   takes the form of a spherical or spheroidal cell   that acts  on the surrounding fluid with a tangential slip velocity prescribed along its surface. The squirmer model has been widely used to study the motion of ciliates such as the green algae \textit{Volvox} and the protozoan \textit{Paramecium} driven by the synchronized metachronal beating of short cilia covering the cell body. A squirmer is also the appropriate mathematical model to   describe the motion of a self-propelled droplet driven by marangoni stresses~\cite[]{thutupalli2011swarming, maass2016swimming} and autophoretic colloidal particles  driven by  electrophoresis, diffusiophoresis or thermophoresis, be they externally driven or self-generated~\cite[]{saville1977electrokinetic, anderson1989colloid, moran2017phoretic}.

In the squirmer model, the tangential velocity is typically taken to be axisymmetric (and therefore with the same symmetry axis as the cell body) and is given by an expansion in  Legendre polynomials (and  associated Legendre polynomials) with either steady or time-periodic coefficients.  In a Newtonian fluid, the flow field around the squirmer is then computed using spherical and spheroidal harmonics~\cite[]{lighthill1952squirming,blake1971spherical,theers2016modeling}. In most applications only the first few modes of the squirming motion, which decay spatially the slowest, are considered;  \textcolor{black}{the influence of higher-order modes on   swimming  in a viscoelastic fluid was considered in \cite[]{datt2017active}.}

In the case of a spherical squirmer, the slip velocity  is often written as  the sum of two modes, $u_\theta^s=B_1\sin\theta+B_2\sin2\theta/2$; the first mode determines the swimming speed ($U_0=2B_1/3$ in an unbounded Newtonian fluid) and represents a source dipole singularity decaying as $u\sim r^{-3}$ in the far-field; the second mode $B_2$ corresponds to a stokeslet dipole decaying as $u\sim r^{-2}$ and a source quadrupole decaying as $u\sim r^{-4}$.  
When $\beta\equiv B_2/B_1>0$, the squirmer is a puller with propulsion generated in front of the body, as in  cells of the green algae genus \textit{Chlamydomonas}; in contrast, when $\beta<0$ the squirmer is a pusher with the propulsion generated from its rear side,  as in the   flagellated bacterium \textit{E.~coli}. The particular case $\beta=0$  corresponds to a neutral squirmer. Note that squirmers with time-periodic modes have also been studied~\cite[]{pedley2016spherical}, as have squirmers with swirls and  azimuthal components to their  boundary conditions~\cite[]{pak2014generalized,pedley2016squirmers}.

In the case $De\ll1$ or $\hat{\mu}_p\ll1$, the motion of a squirmer in an unbounded viscoelastic fluid can be tackled using  asymptotic expansions and the reciprocal theorem. \textcolor{black}{In both cases, the leading-order solution is the squirmer motion in a Newtonian fluid, which induces  the polymer deformation. The difference is that for $De\ll1$, components of the polymer stress tensor are decoupled and are determined solely by the local strain rate, while in the limit $\hat{\mu}_p\ll1$, the components are coupled to one another.}
De Corato et al.~\cite[]{de2015locomotion} considered  spherical squirmers in a second-order fluid in the limit $De\ll1$ and obtained their swimming velocity and power as
\begin{subequations}\label{eq:squirmer}
\begin{equation}
\frac{U}{U_0}=1-\frac{3}{10}\left(1+2\frac{\Psi_2}{\Psi_1}\right)\beta De,
\end{equation}
\begin{equation}
\frac{P}{P_0}=1-\frac{9}{10}\left(1+2\frac{\Psi_2}{\Psi_1}\right)\frac{4\beta+\beta^3}{2+\beta^2}De,
\end{equation}
\end{subequations}
where $U_0=2B_1/3$ and $P_0=6\pi(2+\beta^2)$ are the velocity and power in a Newtonian fluid of the same total viscosity $\mu=\mu_s+\mu_p$ as the viscoelastic fluid, $\Psi_1$ and $\Psi_2$ are the first and second normal stress difference coefficients for the  fluid 
($\Psi_1>0$ and $-\Psi_1/2\leq\Psi_2\leq0$ for a realistic polymeric fluid), and the Deborah number is defined as $De=\lambda U_0/a$ with $a$ the radius of the squirmer and $\lambda=\Psi_1/(2\mu_p)$.

\begin{figure*}[t]
\begin{center}
\includegraphics[angle=0,scale=0.29]{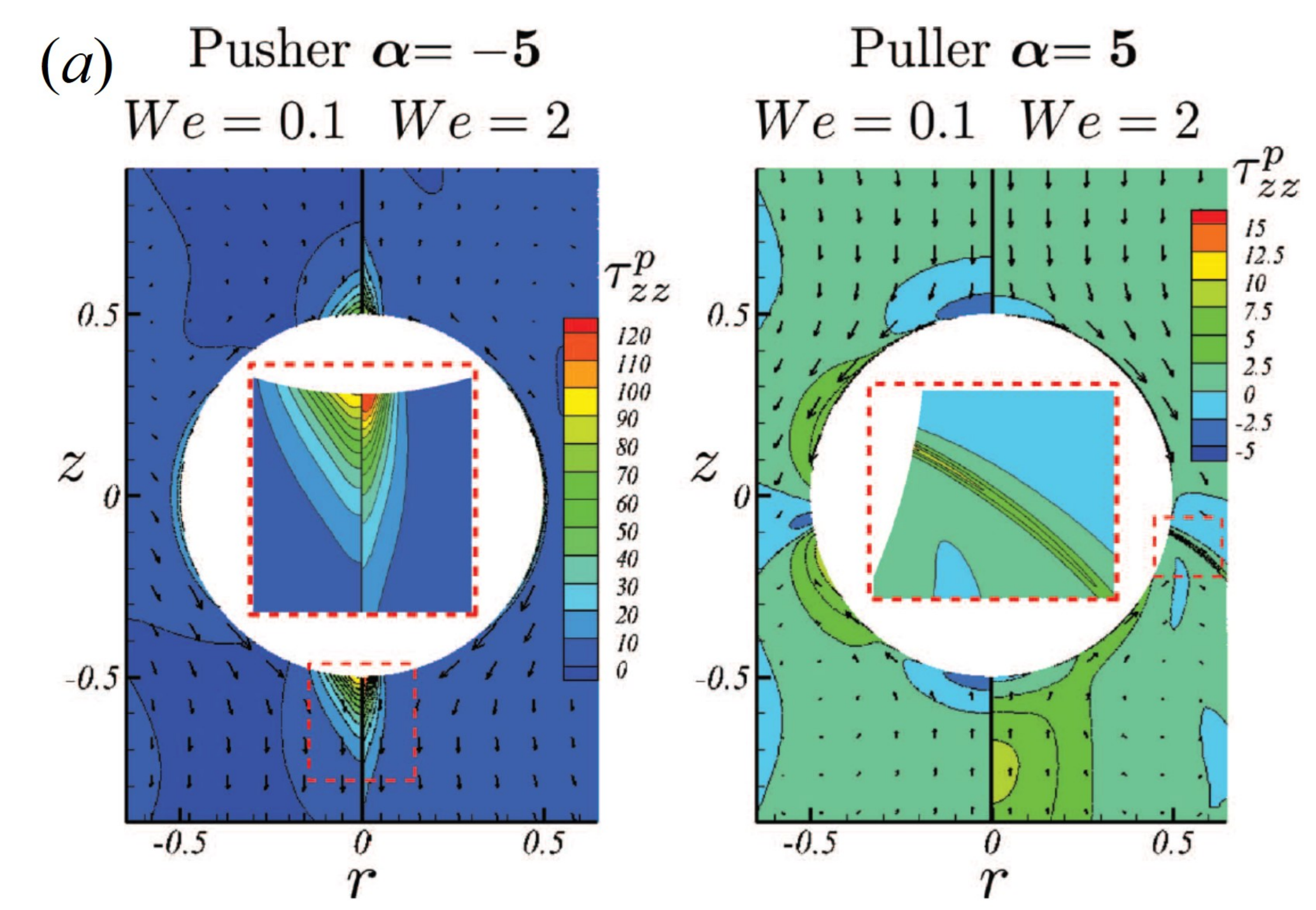}
\includegraphics[angle=0,scale=0.29]{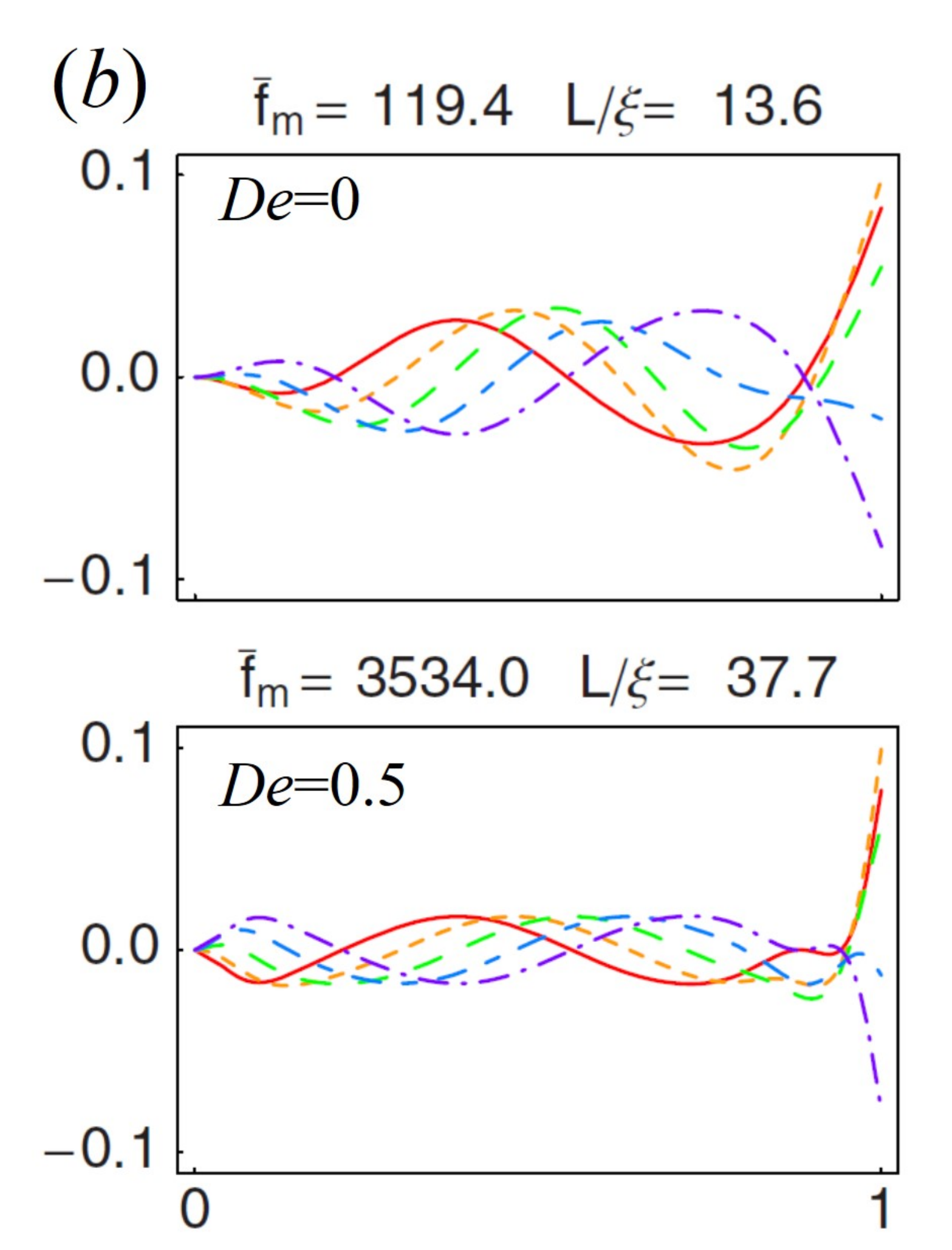}
\includegraphics[angle=0,scale=0.29]{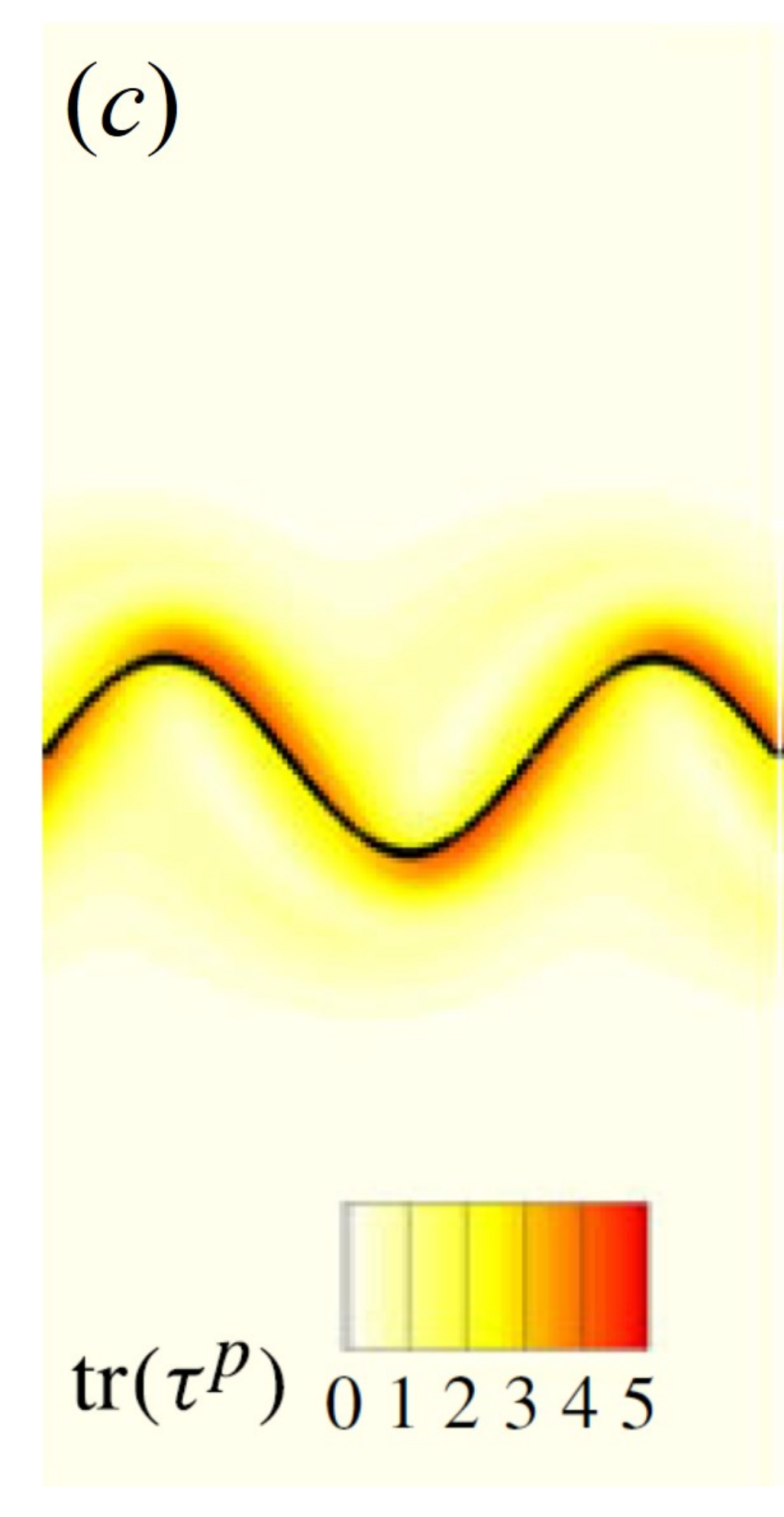}
\includegraphics[angle=0,scale=0.29]{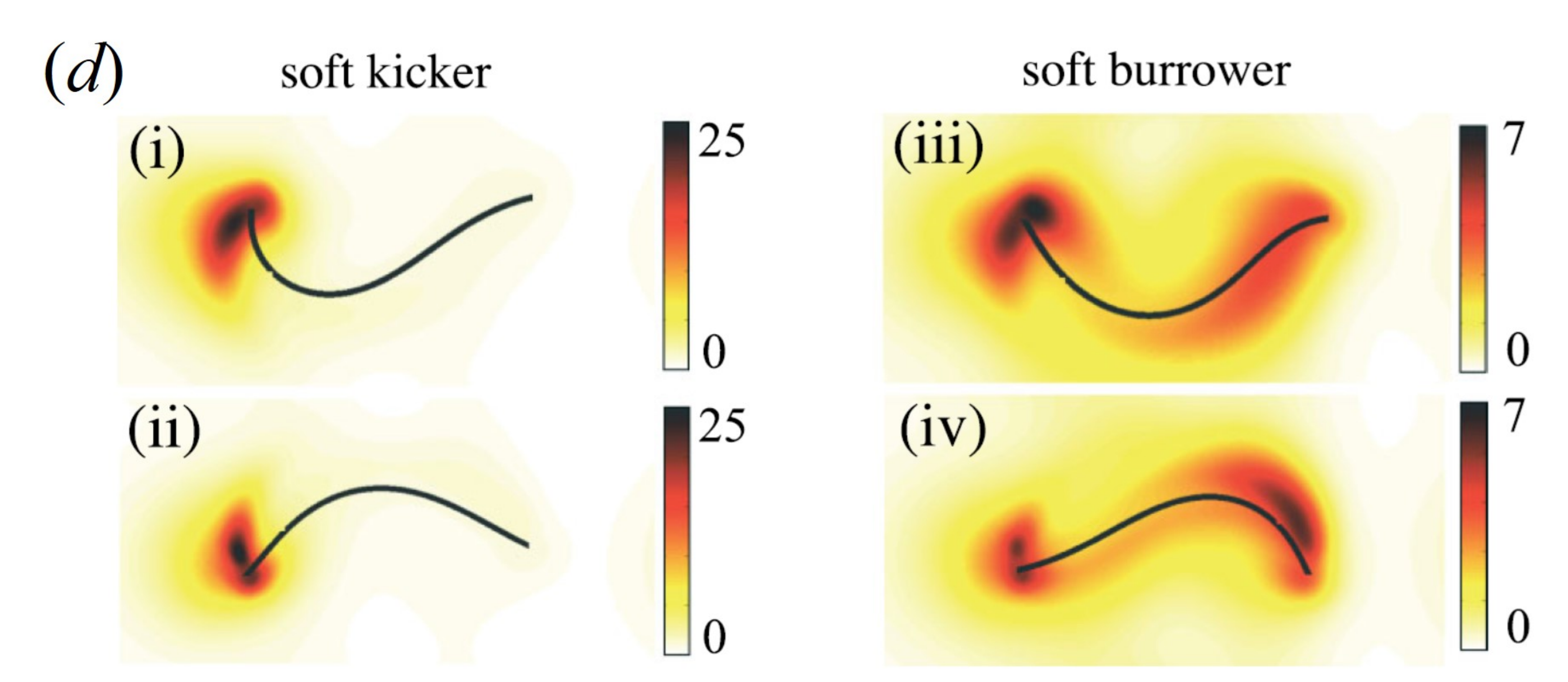}
\includegraphics[angle=0,scale=0.29]{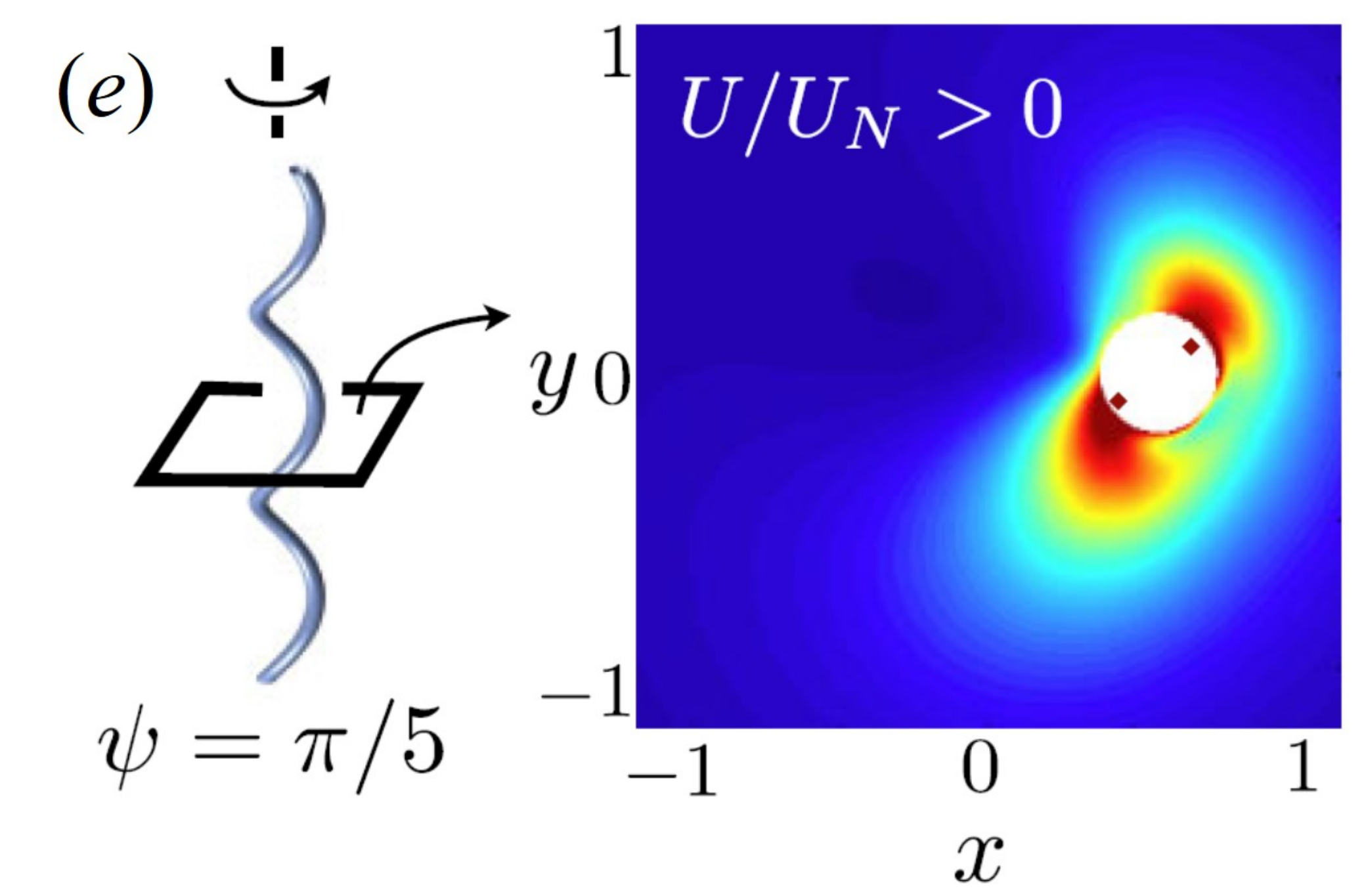}
\caption{Motility of individual model swimmers in viscoelastic fluids. 
($a$) Distribution of polymer stress in   regions of high extensional flow around a squirmer self-propeling upwards; the polymer stresses cause extra resistance and reduce the swimming speed  (adapted with permission from Ref.~\cite[]{zhu2012self}). ($b$) A  flagellum waving in a viscoelastic fluid has a reduced wave amplitude near its proximal end compared to its  Newtonian waveform (adapted with permission from Ref.~\cite[]{fu2008beating}). ($c$) Polymer stress distribution around an infinitely-long two-dimensional sheet  undergoing waving motion  from right to left (adapted with permission from Ref.~\cite[]{li2015undulatory}).
($d$) Distribution of polymer stresses around finite two-dimensional undulatory swimmers self-propeling to the right; the  high polymer stresses at the tail of a soft kicker enhances the swimming speed (adapted with permission from Ref.~\cite[]{thomases2014mechanisms}).
  ($e$) Distribution of polymer stresses around a rotating rigid helix, as a model for a bacterial flagellar filament; at sufficiently high frequencies, the  helical filament revisits its own viscoelastic wake, which enhances its  force-free linear speed    (adapted with permission from Ref.~\cite[]{spagnolie2013locomotion}).
 }\label{fig:speed}
\end{center}
\end{figure*}

This second-order fluid model is valid for a slow and slowly varying flow of a polymeric fluid and most nonlinear constitutive equations reduce to the second-order fluid in the limit $De\ll1$. We have in particular $\Psi_1=2\lambda\mu_p, \Psi_2=0$ for the Oldroyd-B and FENE models, and $\Psi_1=2\lambda\mu_p, \Psi_2=-\lambda\mu_p\alpha_m$ for the Giesekus model (here $\alpha_m$ is the mobility factor). The result in Eq.~(\ref{eq:squirmer}) shows that  pusher cells  ($\beta < 0$)
swim faster but consume more power in a weakly viscoelastic fluid than in a Newtonian fluid; the opposite is true for pullers ($\beta > 0$), that swim slower but expend less power to do so. At leading-order in $De$, neutral squirmers are not affected by the viscoelastic fluid.  
Numerical simulation with a Giesekus model show that the above asymptotic solution remains valid as long as $De<0.02$ and the squirmer speed is always reduced for $De\gtrsim0.1$~\cite[]{de2015locomotion}. Expansion at higher orders in $De$ shows that the neutral squirmer actually swims slower in a Giesekus fluid~\cite[]{datt2019note}.

Similar conclusions appear to apply qualitatively  to autophoretic Janus particle~\cite[]{natale2017autophoretic, Choudhary2020nonnewtonian}. In that case, the type of swimmer (pusher vs.~puller) is determined by the sign of the local  phoretic mobility   and the ratio between active and inert regions.  
By expanding in the viscosity ratio $\hat{\mu}_p$, Li and Koch~\cite[]{li2020electrophoresis} considered the motion of an electrophoretic particle in a dilute polymer solution at finite $De$. While in a Newtonian fluid, a uniformly-charged particle with a thin ionic Debye layer disturbs the fluid as a neutral squirmer, in a polymer solution the particle becomes a puller at small $De$. The particle velocity is reduced for all $De$, with a minimum   occurring at $De=O(1)$.  
At high $De$, the polymer is strongly stretched by the local extensional flow at the rear side of the particle in both Debye layer and bulk region forming a birefringent strand. These results indicate that the swimming kinematics of a swimmer is strongly influenced by the polymer.

At high Deborah numbers, numerical simulations are needed to capture the full impact of   viscoelasticity for large deformations. Zhu et al.~\cite[]{zhu2012self, zhu2011locomotion} studied the motion of a squirmer with both spherical and prolate spheroidal shapes in a polymeric fluid using  finite elements. For all types of squirmers, the  viscoelasticity in the fluid always reduces both the swimming speed and the power consumption. The squirmer reaches a minimum velocity at $De=O(1)$, at which point the swimming efficiency $\sim U^2/P$ peaks. As expected, the polymers are strongly stretched in regions of local extensional flows close to the squirmer surface,  at the front and the rear in the case of pushers and  on the sides in the case of  pullers (figure~\ref{fig:speed}$a$).

\subsection{Waving of slender bodies and  flexible flagella}

For   undulatory swimmers, both  enhancement and decrease of locomotion speeds have been  observed in viscoelastic fluids depending on the details of the swimming gaits and the fluid rheology. The nematode \textit{C.~elegans} always swims slower in a semi-dilute solution of high-molecular-weight polymer (carboxymethyl cellulose) than in a Newtonian fluid of the same viscosity~\cite[]{shen2011undulatory}. In contrast, in a polymer solution of xanthan gum, \textit{C.~elegans} swims slower in a semi-dilute solution but faster in a concentrated solution of entangled polymer network~\cite[]{gagnon2013undulatory}. 
 
Theoretically, in what was historically the first  paper in the field of low-Reynolds number swimming,  Taylor   proposed to use   a two-dimensional  (2D) waving sheet as a model for the propagation of sinusoidal wave on flexible spermatozoa flagella~\cite[]{taylor1951analysis} and later extended it to the case of three-dimensional (3D)  filaments~\cite[]{taylor1952action}. The sheet model also allows to capture the  waving motion of the envelope of  cilia tips   for large ciliated organisms~\cite[]{childress81,laugabook}. 
Both the sheet and filament models predict, at  small amplitude, the same swimming speed in a Newtonian fluid,  $U_0=\omega k A^2/2$ , where $\omega, k$ and $A$ are the frequency, wavenumber and the amplitude of the waving motion (the small amplitude limit corresponds to the assumption $Ak\ll 1$).

The results for both 2D and 3D undulatory swimmers in a viscoelastic fluid are  in broad agreement.  Early studies showed that the speed of a small-amplitude waving slender swimmer  is unchanged in a linear viscoelastic fluid~\cite[]{fulford1998swimming} and in a second-order fluid at $Re=0$~\cite[]{chaudhury1979swimming}; adding the  inertia from the fluid increases the speed if $Re$ is below a critical value, while the    speed  decreases at large $Re$~\cite[]{chaudhury1979swimming}. However as pointed out in several studies~\cite[]{sturges1981motion, lauga2007propulsion, fu2009swimming}, geometrically nonlinear viscoelastic effects  always need to be included  even for a small-amplitude swimmer: \textcolor{black}{a constitutive relationship needs to be objective, and the transport of  viscoelastic stresses naturally leads to geometrically nonlinear terms in the constitutive modelling~\cite[]{bird1987dynamics}.}

In that case, Sturges~\cite[]{sturges1981motion} \textcolor{black}{showed that the time derivative of the Rivlin-Ericksen tensor at each degree is of the same order as the swimming amplitude, therefore the polymeric stress cannot be simplified using a linear model. Based on an integral constitutive equation, the analysis shows that, at a fixed $El$,} the speed of a waving sheet in a nonlinear viscoelastic fluid  decreases monotonically with increasing $Re$. \textcolor{black}{For a fixed value of $Re$, their result shows that the speed decreases monotonically with $De$, consistent with the result in Eq.~(\ref{eq:flagellum}) for a flagellum  in the absence of inertia.}

 In a classical Oldroyd-B polymeric fluid, the swimming speed of both the small-amplitude waving sheet and waving  filament models are~\cite[]{lauga2007propulsion, fu2007theory, fu2009swimming},
\begin{equation}\label{eq:flagellum}
\frac{U}{U_0}
=\frac{1+\hat{\mu}_sDe^2}{1+De^2},
\end{equation}
where $De=\lambda\omega$ is defined using the undulatory frequency $\omega$ of the sheet and where $U_0$ is the swimming speed in a Newtonian fluid. Importantly,  this result arises from the (geometrically) nonlinear response of the viscoelastic fluid to the small amplitude waving motion since in a linear viscoelastic fluid the speed of the flagellum is unchanged~\cite[]{fulford1998swimming}.  The power $P/P_0$ and the swimming efficiency $\sim U^2/P$ follow the same dependence on $De$ as in Eq.~(\ref{eq:flagellum}) if the Newtonian fluid chosen for comparison is taken to have  \textcolor{black}{the  zero-shear rate viscosity of the non-Newtonian fluid.}

The result in Eq.~(\ref{eq:flagellum}) turns out to be  valid for waves of either tangential and normal motion~\cite[]{lauga2007propulsion}, for  filaments undergoing small-amplitude helical waves~\cite[]{li2015swimming}, and for pumping using 
small-amplitude peristaltic  waves (\textcolor{black}{for which the pumping  rate is analogous to the swimming speed})~\cite[]{bohme1983peristaltic, bohme2013analysis}. A squirmer driven by  small-amplitude oscillating slip velocities also shows a similar speed dependence on $De$ in a viscoelastic fluid~\cite[]{lauga2009life, lauga2014locomotion}.  Interestingly,  the speed ratio in Eq.~\eqref{eq:flagellum} is exactly the same as the dimensionless dynamic viscosity $\mu'(\omega)/\mu$ (i.e.~the real part of the complex viscosity) in a small-amplitude oscillatory shear (SAOS) flow~\cite[]{bird1987dynamics, morrison2001understanding}.  The result in Eq.~(\ref{eq:flagellum}) was also found to be independent of the particular constitutive model chosen for the polymeric fluid as along as they have the same complex viscosity as those of Oldroyd-B, Johnson-Segalman, Giesekus and FENE-P~\cite[]{lauga2007propulsion}.

Extensions of the result in Eq.~(\ref{eq:flagellum}) to higher orders in the waving amplitude shows that that the swimming speed of a large-amplitude waving sheet also monotonically decreases with  $De$~\cite[]{elfring2015theory}. The property that the propulsion speed depends nonlinearly on the wave frequency in a complex fluid can be also exploited to design a swimmer whose speed increases in a viscoelastic fluid by imposing multiple waves propagating with different frequencies and amplitudes  in opposite directions~\cite[]{riley2015small}. It was also further shown that the speed change in a viscoelastic fluid depends strongly on the swimming gait, and a speed enhancement is possible for a sheet of unidirectional traveling waves with both transverse and longitudinal deformation modes~\cite{elfring2016effect}.

 \def\Sp{\rm Sp}
A different kind of viscoelastic swimming enhancement  is   possible for a waving flagellum when the feedback from the fluid stresses on the  internal actuation, and therefore the waving amplitude, is taken into account~\cite[]{riley2014enhanced}.  In the simplest case of a small-amplitude waving sheet, the waving motion  can be modeled as driven by a time-varying distribution of active bending forces, and its shape is determined by the solution to the fluid-structure interaction problem, 
i.e.~the instantaneous mechanical balance between the internal actuation, the external hydrodynamic stresses and the passive elastic resistance of the swimmer.  The key dimensionless parameter for an elastic swimmer is called the Sperm 
number and defined as
\begin{equation}\label{eq:sp}
\Sp=(\omega\zeta/\kappa k^n)^{1/n},
\end{equation}
where $\kappa$ is the bending stiffness of the swimmer body, \textcolor{black}{$k$ is the wavenumber,} $n=3$ or 4 for 2D/3D problems, and $\zeta$ is the viscous friction coefficient proportional to the viscosity of the fluid $\mu$. Physically, the  Sperm number quantifies the ratio of viscous to elastic (bending) stresses so the swimmer is relatively soft when $\Sp>1$ but stiff when $\Sp<1$.  
 It can also  be viewed as the ratio between the wavelength  of the waving motion  and the bending length scale, or as  (the $n$th root of) the ratio between the  bending relaxation time of the swimming body and the  period of the waving motion.

For an infinite 2D waving  sheet, the swimming speed is given by~\cite[]{riley2014enhanced}
\begin{equation}
\frac{U}{U_0}=\frac{(1+4{\Sp}^6)(1+\hat{\mu}_sDe^2)}
{1+De^2+4\Sp^3(1-\hat{\mu}_s)De+4\Sp^6(1+\hat{\mu}_s^2De^2)},
\end{equation}
which agrees with the Eq.~(\ref{eq:flagellum}) in the limit $\Sp\ll1$ and becomes $U/U_0=[1+\hat{\mu}_sDe^2]/[1+\hat{\mu}_s^2De^2]>1$ for $\Sp\gg1$.  
\textcolor{black}{For a stiff swimmer with  $\Sp\ll1$, the active stress scaling as $\sim fk^2$ ($f$ is the imposed active bending moment per unit length) is balanced by the bending resistance of the body ($\sim\kappa Ak^4$), so the waving amplitude scales as $A\sim f/\kappa k^2$ and it is not affected by the fluid. In that limit, the velocity decreases in a viscoelastic fluid as described in Eq.~(\ref{eq:flagellum}). For $De\gg1$, the speed is $\hat{\mu}_s$ times the speed in a Newtonian fluid ($\hat{\mu}_s$ is the dimensionless solvent viscosity). In contrast, for a soft swimmer with $\Sp\gg1$, the active stress is balanced by the viscous resistance from the fluid. In a Newtonian fluid, that fluid resistance scales as  $\sim\mu U_N/A\sim\mu\omega Ak$, where we have used Taylor's result for swimming speed $U_N\sim A^2\omega k$. In a viscoelastic fluid with $De\gg1$, the fluid resistance $\mu U_V/A_V$ becomes $\mu_s\omega A_Vk$ due to $U_V\sim\hat{\mu}_s U_0$ (Eq.~\ref{eq:flagellum}) and $U_0\sim A_V^2\omega k$ where $A_V$ is the amplitude in a viscoelastic fluid, $\mu$ and $\mu_s=\hat{\mu}_s\mu$ are the total and solvent viscosity, respectively. The two fluid resistances are balanced by the same active stress $\sim fk^2$, therefore $A_V\sim A/\hat{\mu}_s$} \textcolor{black}{and $U_V\sim U_N/\hat{\mu}_s$. For a soft flagellum, the waving amplitude is seen to increase in a viscoelastic fluid, so much so that it may overcome the viscoelastic hindrance from Eq.~(\ref{eq:flagellum}) and gain an overall speed enhancement.}
 
For a swimmer of finite length, the waveform   may no longer be a simple sinusoidal wave. In addition to the effects above, \textcolor{black}{the analysis of a finite filament undergoing small-amplitude waving showed that the interaction between an elastic body and a viscoelastic fluid} can modify the swimming speed by qualitatively affecting  the beating pattern of the filament~\cite[]{fu2007theory, fu2008beating}. An  increase in $De$ tends to reduce the swimming speed monotonically for a filament of large wavelength; for a filament of medium wavelength the speed 
 first slightly increases and then it decreases; finally an increase in $De$ can  reverse the swimming direction of a filament with small wavelength by dramatically changing its beating pattern.

 Physically,  the elasticity of the fluid  increases the bending length   of the filament and tends to suppress its undulations at its proximal end (i.e.~the junction to the  cell body) and in the middle portion (figure~\ref{fig:speed}$b$). The results in Ref.~\cite{fu2008beating}   qualitatively reproduce the  beating patterns of spermatozoa flagella observed in Newtonian and viscoelastic fluids~\cite[]{katz1978movement, ishijima1986flagellar, ho2001hyperactivation}. 
A similar effect can be observed for the flagella on the algae \textit{C.~reinhardtii} whose lateral displacements  near the cell body decrease in a viscoelastic fluid~\cite[]{qin2015flagellar}.

\textcolor{black}{Two-dimensional} numerical simulation  demonstrated that a waving swimmer of finite length and large amplitude can swim faster in a viscoelastic fluid~\cite[]{teran2010viscoelastic}. It has a maximum speed   at $De\sim1$ and generates strong polymer extension behind it.  This result turns out to be very different from the polymer stress distribution around an infinitely long flagellum, for which the polymer is periodically stretched on the windward side of the traveling wave and relaxes on the leeward side (figure~\ref{fig:speed}$c$).  Thomases and Guy~\cite[]{thomases2014mechanisms} further showed that, beyond the fluid,  the swimming speed in a viscoelastic fluid  depends critically on the  elasticity of the  body and the  swimming stroke. Specifically, a swimmer  with a decreasing amplitude from head to tail (e.g., \textit{C.~elegans}) slows down in a viscoelastic fluid, and so does a stiff swimmer whose waving amplitude increases  (e.g.~flagellated spermatozoa). In contrast, if that swimmer is  softer, the swimming speed can  go up  due to the combined effects of the increase in the stroke amplitude~\cite[]{thomases2017role} and the favorable asymmetric distribution of the polymer stress around the swimmer (figure~\ref{fig:speed}$d$). \textcolor{black}{This result, which is consistent with  theoretical results for an infinitely-long elastic swimming sheet internally driven by  active stresses~\cite[]{riley2014enhanced}, can  help explain the experiments in Refs.~\cite[]{shen2011undulatory, gagnon2013undulatory}. In a concentrated polymer solution, the viscosity of the fluid substantially increases, which leads to an increase in the Sperm number from Eq.~(\ref{eq:sp}), or equivalently, a reduction of the effective stiffness of the swimmer.} In three dimensions, the polymer stresses become more concentrated near the end of the swimmer and lead to a weaker speed reduction for a model \textit{C.~elegans} \textcolor{black}{with prescribed kinematics (i.e.~in the infinitely stiff limit)} compared to the 2D case~\cite[]{binagia2019three}. 

Experimentally, an increase in the swimming speed was demonstrated in a viscoelastic fluid for  a flexible swimmer  composed of a magnetic head and a soft tail actuated in a time-periodic magnetic field~\cite[]{espinosa2013fluid}. 
Within the range of parameters tested in the experiments ($De\leq5$), the  ratio of swimming speed for swimmers in Boger  fluids (i.e.~viscoelastic fluids that have a constant viscosity) and Newtonian fluids of the same viscosity was seen to continuously increase with the Deborah number. Related work showed that  speed of a waving cylindrical sheet increases in a Boger fluid, while it decreases in a shear-thinning viscoelastic fluid~\cite[]{dasgupta2013speed}, showing that the shear-dependence of the viscosity  can also have  a strong impact on microswimming.

\subsection{Helical locomotion}

Microswimming driven by rotating slender helical shapes is the locomotion method used by bacteria with slender flagellar filaments and by spirochetes. Early experiments showed that many bacteria can not only swim  in polymer solutions~\cite[]{schneider1974effect, kaiser1975enhanced, greenberg1977relationship, greenberg1977motility, berg1979movement} but that in fact  they swim faster than in the solvent  alone,  despite the fact that the polymer solution  is much more viscous~\cite[]{kaiser1975enhanced}.  
For a wide range of bacteria species, the average speed  varies non-monotonically with the solution viscosity and reaches a peak value at a critical viscosity~\cite[]{schneider1974effect, greenberg1977relationship, greenberg1977motility}.

Berg and Turner~\cite[]{berg1979movement} showed that the angular velocity of the body of a tethered \textit{E.~coli} cell is inversely proportional to the fluid viscosity in a   branched polymer solution, consistent with  Newtonian dynamics; in contrast, it is less affected in a solution of   unbranched polymer chains, suggesting strong non-Newtonian effects in that case.  Berg and Turner suggested that in a solution of unbranched polymers, bacteria swim in a loose quasi-rigid network of microsize pores and experience as a result two different resistances when moving through the network (low friction inside the pores and high friction against the network), a strong anisotropy at the origin of the unusual swimming behavior. In comparison, a solution of   branched polymers is more homogeneous and has a behaviour closer to that of a Newtonian fluid. 
Based on this physical idea, Magariyama and Kudo~\cite[]{magariyama2002mathematical} modified the classical resistive-force theory empirically by introducing different apparent viscosities for motions tangent and normal to a body surface and successfully reproduced experimental results.

A recent study~\cite[]{martinez2014flagellated} reexamined the Berg and Turner argument and  demonstrated convincingly that the rotating rate of the cell body seen in a solution of unbranched polymers  originates from a  viscosity contrast between that experienced by the flagellum and that near the  body. Indeed, around the fast-rotating flagellum, the transition of polymers from a coiled to an elongated state creates locally a low viscosity zone and results in an  enhancement of the swimming  speed~\cite[]{martinez2014flagellated}.  
This study  highlighted therefore that the swimming dynamics is strongly influenced by the microrheology  of the non-Newtonian fluid rather than the properties at macroscales assumed in a classical continuum fluid model.  We will revisit this  idea in the next subsection and focus first below on studies that modeled  the viscoelastic fluid as a continuum material.
 
\textcolor{black}{For an infinitely-long filament passing a helical travelling wave of small amplitude}, the swimming speed in a continuum viscoelastic fluid monotonically decreases with increasing $De$ and follows exactly the same expression (\ref{eq:flagellum}) as a waving sheet and filament~\cite[]{fu2009swimming}.  By solving the Stokes equations in a helical coordinate system, Li and Spagnolie~\cite[]{li2015swimming} studied the locomotion of helical bodies of arbitrary cross-section using helical waves.  They showed that, in a viscoelastic fluid and for helices of arbitrary cross-section,  the speed of each mode of the small-amplitude helical wave is proportional to the real part of the complex viscosity, a result that remains true for   swimming confined inside a cylindrical tube.

In contrast, a helical filament  with a large pitch angle was shown to swim with a speed  that depends non-monotonically on the Deborah number~\cite[]{liu2011force, spagnolie2013locomotion}.  Experiments demonstrated that macro-scale force-free rigid helices  rotating along their axes in viscous and viscoelastic Boger fluids experience a  maximum speed enhancement   for $De\sim\textrm{O}(1)$, which was confirmed by numerical simulations~\cite[]{spagnolie2013locomotion}.  At the same Deborah number, the speed enhancement was found to be stronger for a helix of high pitch angle and small  thickness than for low pitch angle and large radius. 
 Using simulations to solve for the distribution of polymer stresses,  the speed increase  was suggested to be caused by the interaction of the rotating helical shapes with their own viscoelastic wakes, which is   strongest when the time scale of the helical motion matches the polymer relaxation time (figure~\ref{fig:speed}$e$)~\cite[]{spagnolie2013locomotion}.

\begin{figure*}[t]
\begin{center}
\includegraphics[angle=0,scale=0.31]{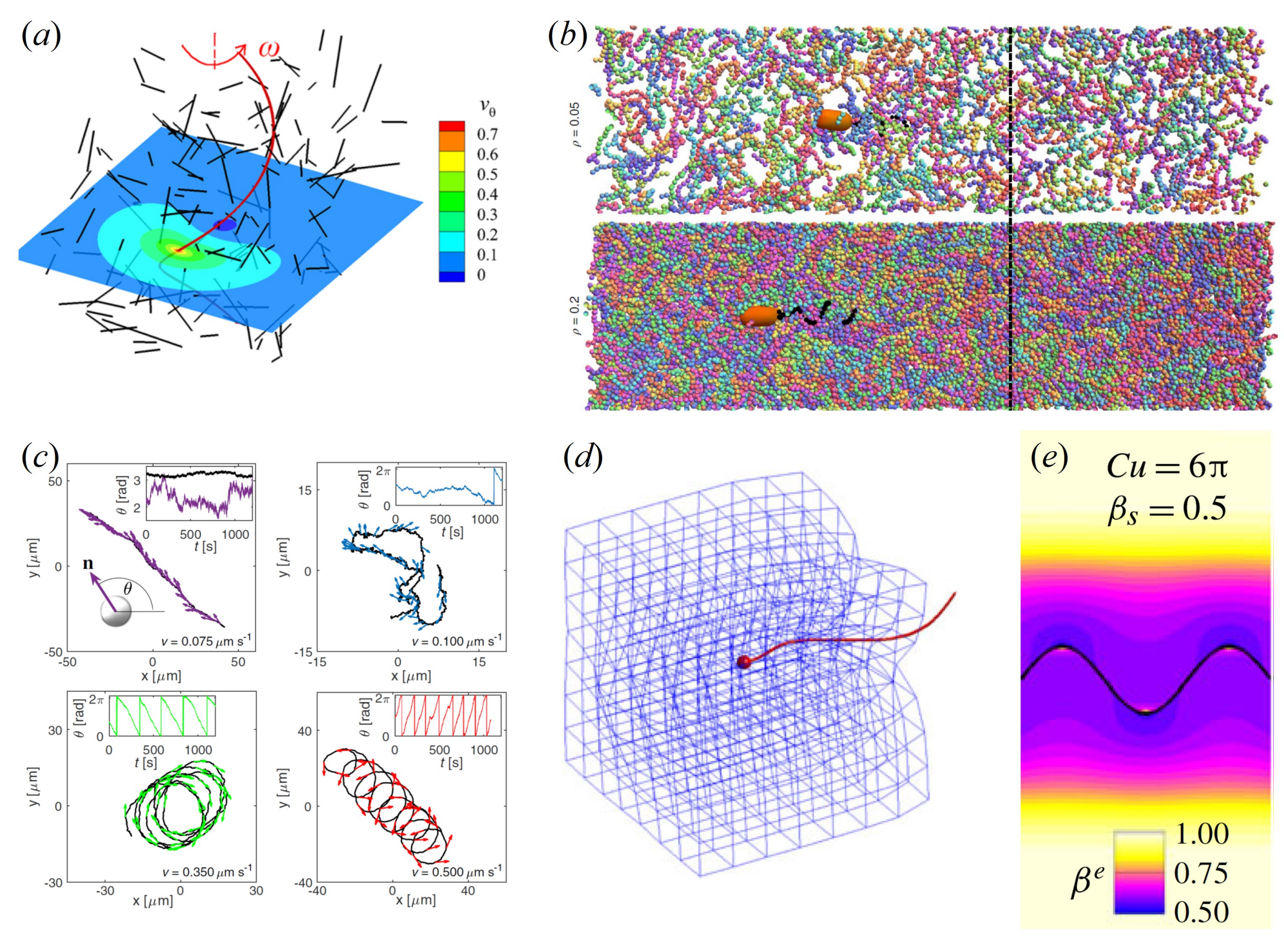}
\caption{Solving for the microstructure of complex fluids during cell swimming. ($a$) Distribution of long polymer dumbbells around a rotating helical filament (adapted with permission from Ref.~\cite[]{zhang2018reduced}). 
($b$) Bacteria swim faster in a dense solution  of semi-flexible polymers than in a dilute solution (adapted with permission from Ref.~\cite[]{zottl2019enhanced}). 
($c$) In a polymer solution, increasing the  speed of an active Janus particle first increases the angular mobility of the particle and then turns its trajectory into circles (adapted with permission from Ref.~\cite[]{narinder2018memory}). 
($d$) Model flagellated swimmer moving through a viscoelastic network (adapted with permission from Ref.~\cite[]{wrobel2016enhanced}). 
($e$) An infinitely long waving flagellum swims faster in a shear-thinning fluid than a Newtonian one  due to an effective confinement effect (adapted with permission from Ref.~\cite[]{li2015undulatory}).
}\label{fig:othereffects}
\end{center}
\end{figure*}

\textcolor{black}{\subsection{Non-continuum effects}}

As seen in this review so far, most studies on viscoelastic microswimming   have modeled the fluid as single-phase continuum medium. However, complex fluids are usually chemically and mechanically heterogeneous on the length scales relevant to biological swimmers, so non-continuum effects  are therefore expected to be important.  

In dilute and semi-dilute polymer solutions, microswimmers see individual and overlapping polymer coils of radius of gyration $\sim10-10^2$~nm. The  rotating flagellar filaments of bacteria induce a local shear flow which strongly stretches polymers~\cite[]{zhang2018reduced,patteson2015running},  reducing the fluid viscosity around the flagellum. This in turn increases the cell swimming speed by increasing the flagellum rotation and decreasing the cell body rotation~\cite[]{martinez2014flagellated,zhang2018reduced}.

 Brownian dynamics simulations for a helix rotating in a suspension of polymer dumbbells  show that, \textcolor{black}{at the same Deborah number and solvent and polymer viscosities}, long polymers are less deformed than short polymers since they experience a weaker nonlocal shear (figure~\ref{fig:othereffects}$a$)~\cite[]{zhang2018reduced}. 
  The magnitude of  first normal stress difference exerted on the helix,  and the modification of the swimmer speed, are also found to be more pronounced when the   polymer molecule is smaller than the swimmer~\cite[]{zhang2018reduced}.

 By tracking the motion of a single polymer chain near a rotating helix, Balin et al.~\cite[]{balin2017biopolymer} showed that  polymers are pumped along the helical flagellum, while migrating radially towards the rotating helix and being stretched, eventually depositing at the rear end of the flagellum and relaxing to their equilibrium configuration.  When attached to a cell body, the rotating helix creates a depletion zone of long polymer chains leading to an apparent slip velocity between the swimmer and the fluid, which increases the swimming speed~\cite[]{zottl2019enhanced} (figure~\ref{fig:othereffects}$b$). 
Note that  apparent slip was proposed in  earlier theoretical models as a mechanism to increase   swimming speeds~\cite{man2015phase}.
\textcolor{black}{The study in Ref.~\cite[]{zottl2019enhanced}} then predicts that the flagellum rotates slower in a polymer solution than in a Newtonian fluid, which is the opposite of the argument based on the viscosity difference for cell body and flagellum from Ref.~\cite[]{martinez2014flagellated}, and  future experiments will be necessary to measure the angular speed of the flagellum to test these hypotheses.

Beyond flagellated cells,  active Brownian \textcolor{black}{colloidal} particles  are significantly influenced by  polymers in suspension.  Experiments show that, by increasing the Deborah number, 
the rotational diffusivities of the particles increase monotonically by more than two orders of magnitude and saturates at $De\sim0.1$, and the translational diffusivities increase by one order of magnitude~\cite[]{gomez2016dynamics}. Above a critical Deborah number $De>De_c\sim\hat{\mu}_p^{-1}$, active particles undergo a transition from enhanced diffusion to a persistent rotational motion~\cite[]{narinder2018memory} (figure~\ref{fig:othereffects}$c$).

Unlike in a Newtonian environment,  the relaxation of polymers  means that  \textcolor{black}{the flow induced by an active particle in a viscoelastic fluid is influenced by its time history}. Such a fluctuating flow field exerts in turn \textcolor{black}{long-memory random stresses} on the particle and interacts with the Brownian fluctuations. Depending on how the persistence time compares  to the time scale of the particle motion, this can enhance the particle diffusion or even lead to its circular motion~\cite[]{narinder2018memory}.  Coarse-grained simulations further showed that the rotational enhancement of active particles is affected by two other effects, namely the reduced absorption of polymers on the particle surface and a \textcolor{black}{front-back} asymmetric encounter of polymers at the front of the swimmer~\cite[]{qi2020enhanced}. The second effect is more pronounced when the polymer concentration is near the overlap concentration, beyond which the rotational diffusivity is reduced because of strong and homogeneous polymer absorption~\cite[]{qi2020enhanced}.  
\textcolor{black}{In the entangled polymer regime, recent work showed that   active particles can escape   entanglements on time scales much shorter than the polymer relaxation time~\cite[]{saad2019diffusiophoresis}. The competition between  caging in  polymer solutions and the persistent motion of   active particles leads to a non-monotonic dependence of the translational diffusivity on the particle radius, or the inverse of $De$, in a viscoelastic fluid~\cite[]{du2019study}.
} 


\textcolor{black}{\subsection{Effects of polymer entanglement and shear-thinning viscosity}}

At high concentrations, because of crosslinking, polymer molecules might overlap and entangle with one another, thereby forming gel-like networks. Leshansky~\cite[]{leshansky2009enhanced} showed that, in a fluid consist of fixed microstructures, the strong anisotropy in hydrodynamic resistance enhances the speed and efficiency of microswimming with prescribed propulsion gaits.  Studies on  two-fluid models that treat the  polymer network and solvent as two coupled elastic and viscous continuum phases further showed that   microswimming speed is   influenced by the deformability of the network and the nature of the interactions between the swimmer and the network~\cite[]{fu2010low, lee2016immersed}.  
The strongest speed enhancement is observed for a swimmer in a sparse network of stiff polymers when the swimmer/polymer interaction is mediated by the solvent~\cite[]{fu2010low}.  A similar conclusion was obtained by simulations of a discrete model of a viscoelastic network immersed in a viscous fluid~\cite[]{wrobel2016enhanced} (figure~\ref{fig:othereffects}$d$).

Besides   elasticity, the shear-thinning behaviour of the viscosity is another important property for complex fluids. The impact of shear-dependent rheology on microswimming has been studied in previous studies and we refer to them for further details~\cite[]{gagnon2014undulatory, li2015undulatory, datt2015squirming, gagnon2016cost, riley2017empirical, gomez2017helical}. \textcolor{black}{Typically, wherever high shear is induced, the  fluid viscosity decreases locally (figure~\ref{fig:othereffects}$e$) and  causes two effects: a local confinement of the swimmer which enhances the swimming speed~\cite[]{li2015undulatory, gomez2017helical} and reduction of fluid forces which tends to hinder locomotion~\cite[]{montenegro2013physics, datt2015squirming, riley2017empirical}. The power consumption is however always reduced due to the reduction in viscosity~\cite[]{gagnon2016cost}.}

\section{Hydrodynamic interactions in viscoelastic fluids}
\label{sec:int}

In the previous section, we discussed the motion of a single swimmer in an  unbounded viscoelastic fluid, focusing on the change in the swimming speed and energetics. Beyond individual locomotion, other physical effects are impacted by  viscoelasticity, including the response to external flows \cite{kumar2019effect,kumar2019flow,elfring2017force,de2017dynamics,mathijssen2016upstream},  the interactions with surfaces and interfaces \cite{chrispell2013actuated,li2014effect,karimi2015interplay,li2017near,ives2017mechanism,yazdi2012bacterial,yazdi2014locomotion,yazdi2015swimming,narinder2019active}, \textcolor{black}{swimmer-swimmer interactions~\cite[]{chrispell2013actuated,woolley2009study,elfring2010two}} and collective motion~\cite[]{bozorgi2011effect, bozorgi2013role, bozorgi2014effects, li2016collective, ArdekaniDesai2017modeling,hemingway2015active,hemingway2016viscoelastic,emmanuel2020active}.  
These can all significantly influence the behaviour of microorganisms  in nature.  For example, biofilm formation is usually initiated by the accumulation and the irreversible attachment of bacteria on  surfaces~\cite{conrad2018confined}. 
External flows can enhance   bacteria aggregation, trigger biofilm formation in   vortical regions~\cite[]{yazdi2012bacterial} and lead to the formation of biofilm streamers in confined environments~\cite[]{drescher2013biofilm}. Interaction with clean and contaminated interfaces can affect the spatial distribution of bacteria \cite{desai2018hydrodynamics, ahmadzadegan2019hydrodynamic, desai2020biofilms, shaik2019swimming}.
In contrast to the extensive literature on  individual swimming in a bulk viscoelastic fluid, fewer studies have been devoted to aspects of interactions. In this section, we summarize work on the three types of relevant hydrodynamic interactions, namely swimmers with surfaces and interfaces, swimmers with external flows and between neighboring   swimmers.

\begin{figure*}[!ht]
\begin{center}
\includegraphics[angle=0,scale=0.3]{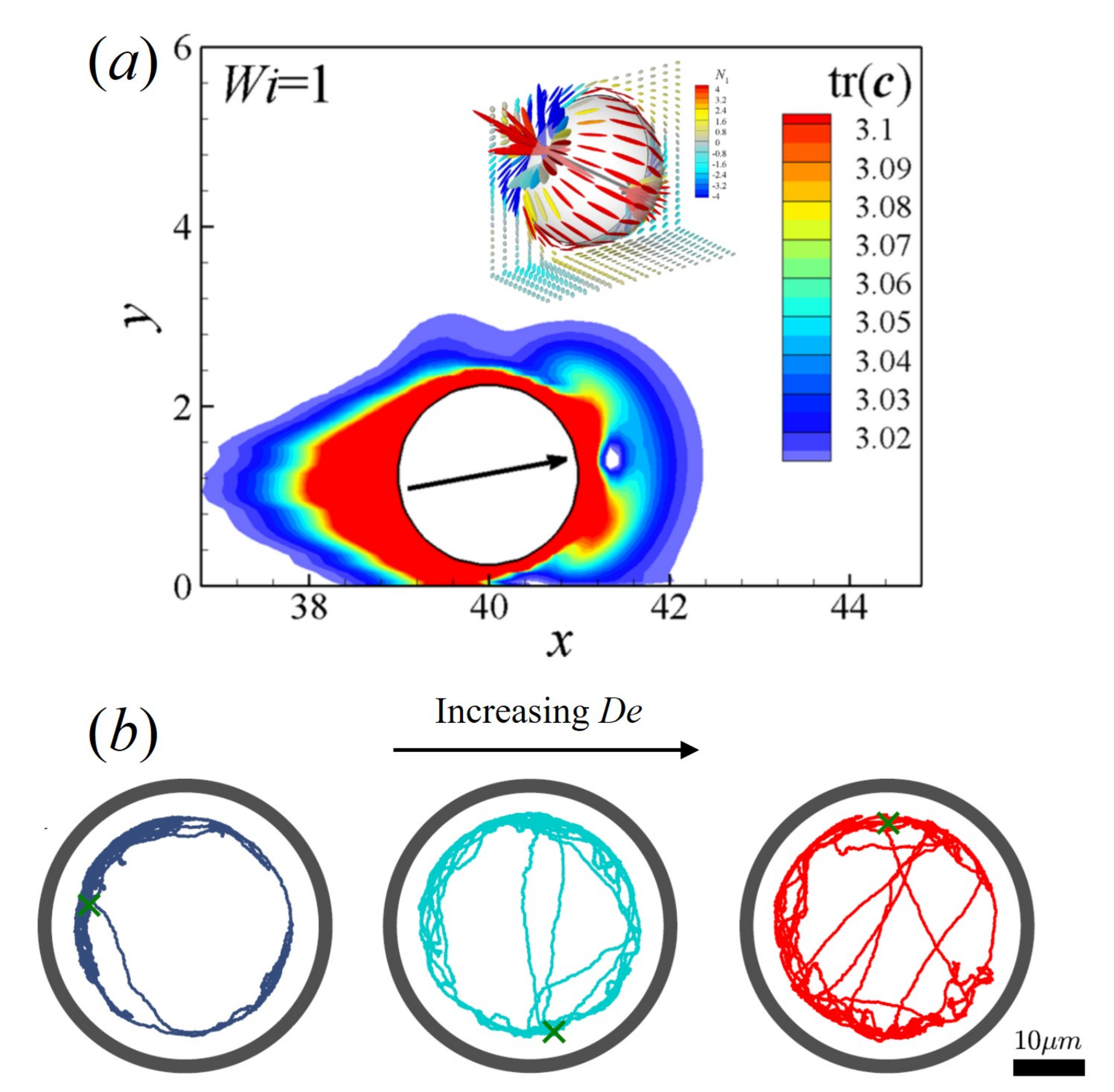}
\includegraphics[angle=0,scale=0.3]{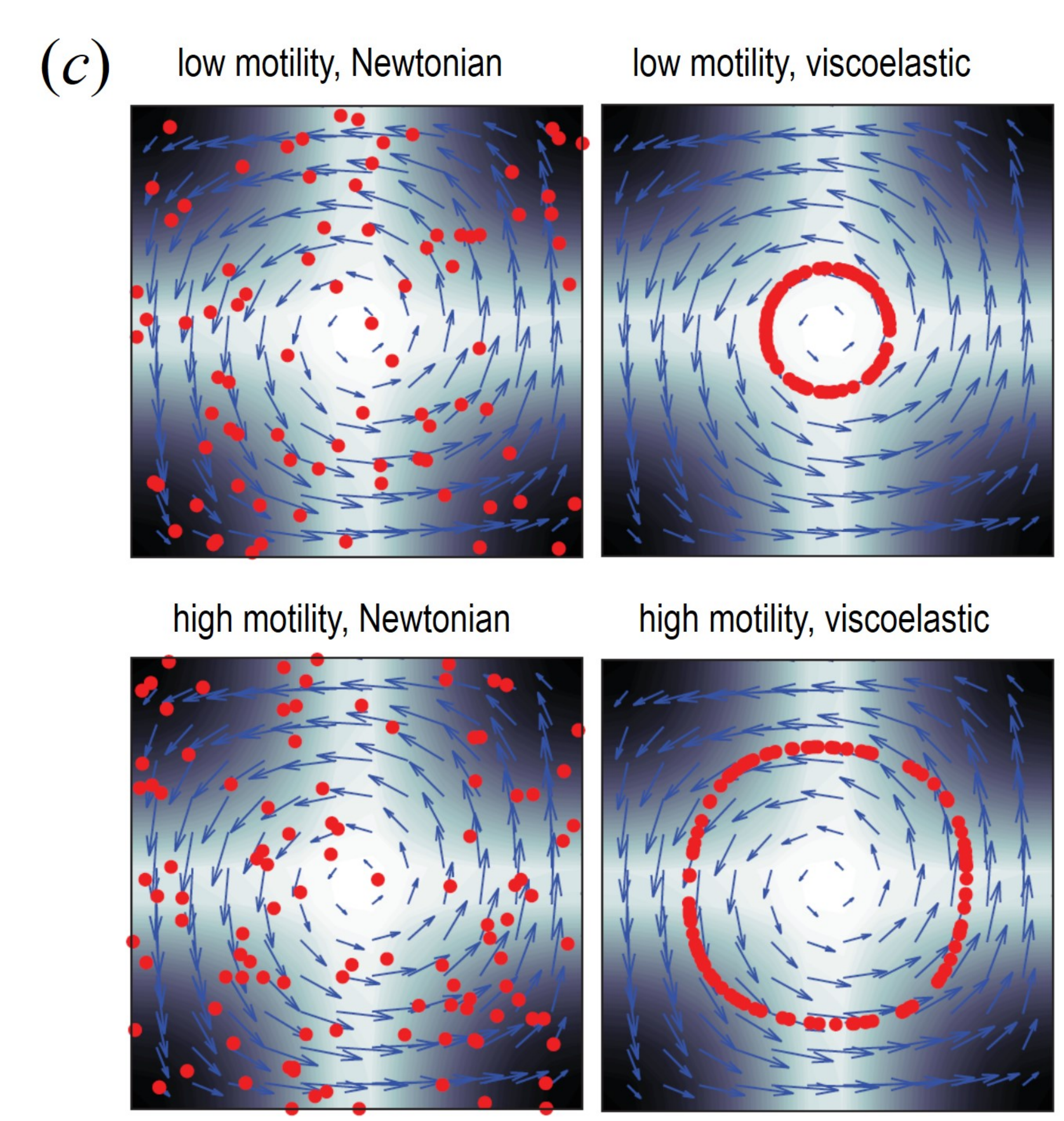}
\includegraphics[angle=0,scale=0.3]{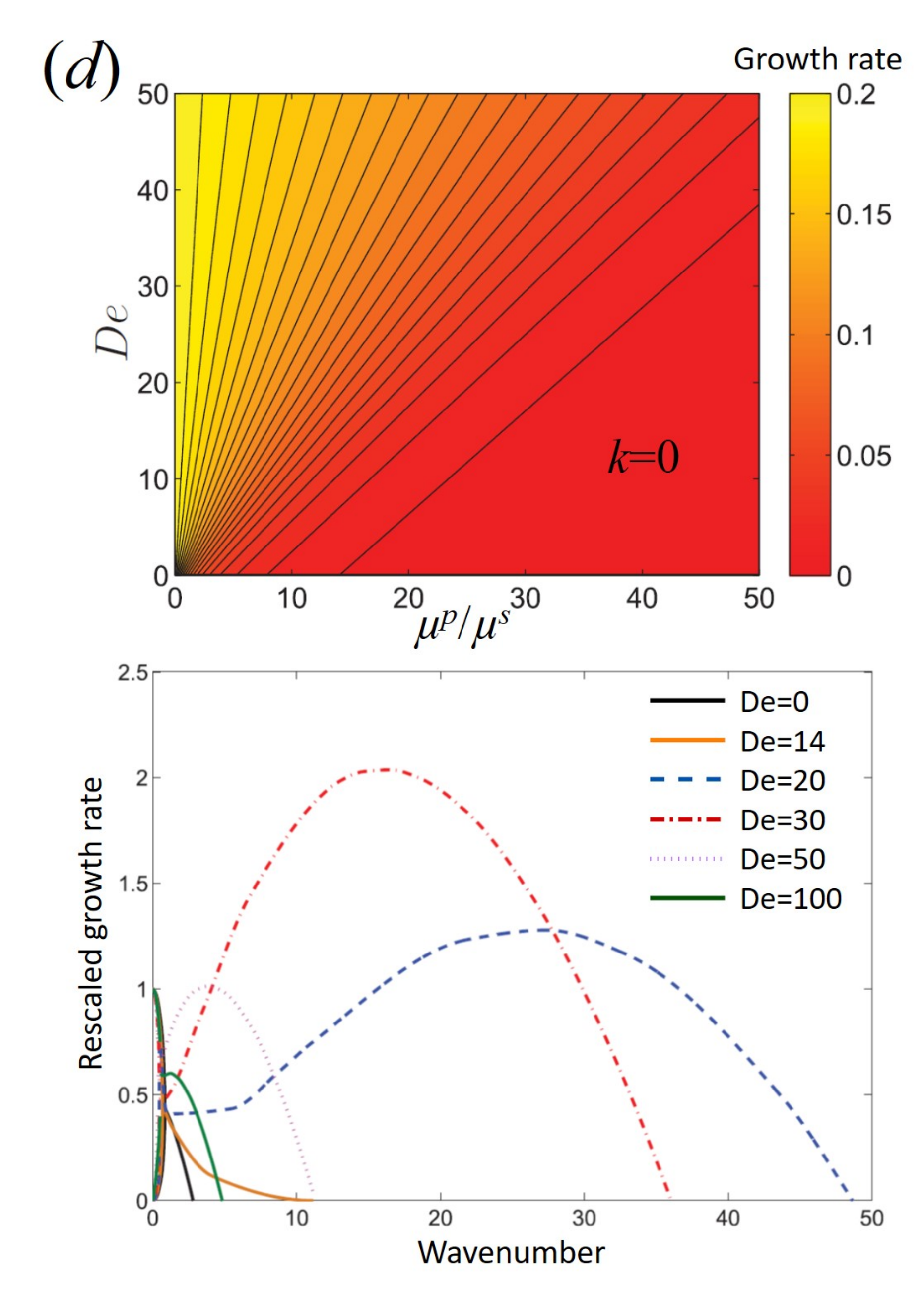}
\includegraphics[angle=0,scale=0.3]{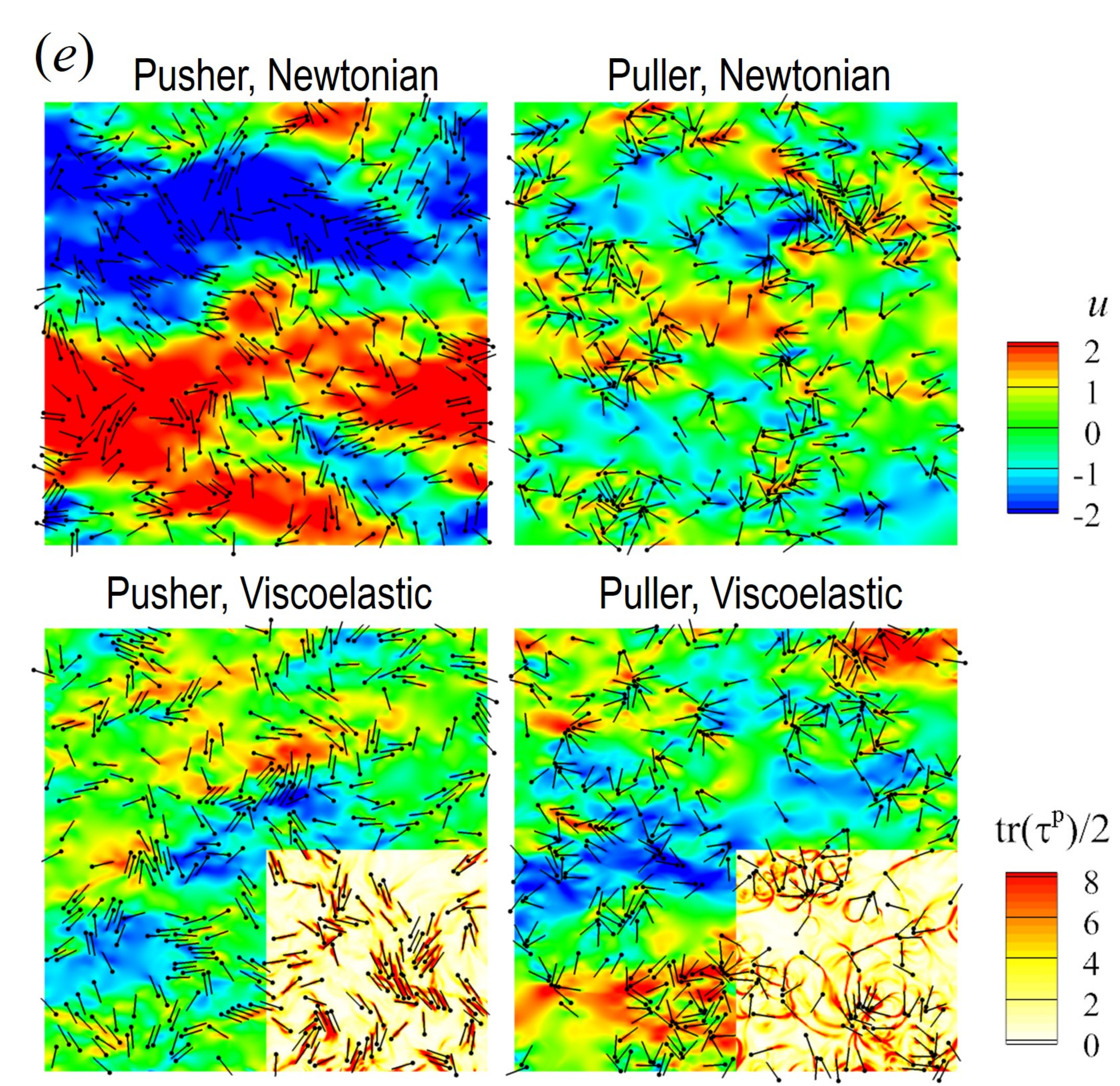}
\caption{Interactions for swimmers in viscoelastic fluids.   
($a$) A pusher squirmer is permanently trapped near a surface due to  large polymer stresses at the tail of the swimmer (adapted with permission from Ref.~\cite[]{li2014effect}). Inset: polymer extension around the squirmer represented by ellipsoids oriented and scaled by the eigenvectors of the polymer conformation tensor. 
($b$) Trajectories of a self-propelled spherical particle (diameter: 7.82 $\mu$m) in a circular domain of a viscoelastic fluid. The particle escapes from the wall more frequently at higher $De$~\cite[]{narinder2019active}. 
($c$) Microswimmers in a vortical flow are randomly distributed if the fluid is Newtonian but they concentrate along limit cycles if the fluid is viscoelastic, with swimmers of higher motility forming a larger cycle (adapted with permission from Ref.~\cite[]{ardekani2012emergence}).
($d$) Growth rate of the   linear perturbation for a homogeneous and isotropic distribution of pushers in a viscoelastic fluid (adapted with permission from Ref.~\cite[]{bozorgi2011effect}).
($e$) Flow field (velocity $u$ and elastic energy $\textrm{tr}(\tau^p)$/2) and distribution of swimmers in Newtonian (top) and viscoelastic fluids (bottom).  Viscoelasticity in the fluid reduces the size and intensity of the coherent flow structure in a pusher suspension and enhances the pusher aggregation, while it has less effects on the pullers (adapted with permission from Ref.~\cite[]{li2016collective}).}
\label{fig:interaction}
\end{center}
\end{figure*}

\subsection{Interactions with surfaces}

Swimming in a viscoelastic fluid near a boundary has been addressed for  Taylor's waving sheet model~\cite[]{chrispell2013actuated, ives2017mechanism, li2017near}. \textcolor{black}{In the small-amplitude limit,  the swimming speed is simply given by the  speed obtained in an unbounded Newtonian fluid, $U_0=\omega kA^2/2$, times two ratios, which are due to two separate effects: a term  identical to that for  near-wall swimming in a Newtonian fluid  (which is always above one and thus enhances swimming)~\cite[]{katz1974propulsion}, 
and the  term from Eq.~(\ref{eq:flagellum}) for  swimming in an unbounded viscoelastic fluid (which is always below one and thus always hinders swimming)~\cite[]{elfring2015theory, ives2017mechanism}. 
At a  finite amplitude,  numerical simulations show that the combination of the wall and viscoelasticity} effects can lead to a non-monotonic speed dependence on $De$ and produce a stronger speed enhancement than the wall effect alone when both the flagellum-wall distance and the normalized  polymer viscosity  are small ($\beta_p=0.1$)~\cite[]{ives2017mechanism}. 
It was conjectured that this speed enhancement is caused by the combination  of a wall-induced increase of     polymer stretching in regions of   extensional flow between the alternating vortices near the waving sheet and the modification of the fluid velocity due to the advection and relaxation of the polymer stress. Chrispell et al.~\cite[]{chrispell2013actuated}  further studied numerically the hydrodynamic interactions between an actuated swimming sheet and \textcolor{black}{a rigid wall or with a passive elastic membrane}, observing a deterioration of the swimming performance in an Oldryod-B fluid compared to a Newtonian one,  \textcolor{black}{resulting from  the relatively large flagellum-surface distance and high polymer viscosity ratio ($\beta_p=0.5$)}.

An important impact of viscoelasticity on  near-wall microswimming is that it can strongly modify a swimmer's trajectory and change its near-wall residence time.  Of course this does not occur for an infinitely-long sheet model since the fluid in the gap is incompressible, \textcolor{black}{and therefore the volume of fluid between the swimmer and the wall cannot change}.  In a Newtonian fluid, it has long been observed  experimentally that  swimming bacteria  are attracted by  surfaces~\cite[]{lauga2006swimming, berke2008hydrodynamic, li2009accumulation, ahmadzadegan2019hydrodynamic}. This   phenomenon results from a combination of multiple effects, including far-field hydrodynamic interactions~\cite[]{berke2008hydrodynamic}, bacteria reorientation due to collisions~\cite[]{drescher2011fluid} or  Brownian fluctuations~\cite[]{li2009accumulation, li2011accumulation}, and surface-induced suppression of flagellar unbundling~\cite[]{molaei2014failed}.

  When only hydrodynamic interactions are included, numerical simulations of squirmer models in a Newtonian fluid show that the squirmers of relatively weak dipoles ($|\beta|\lesssim2$) tend to swim parallel to \textcolor{black}{a no-slip surface during a finite time that depends on many factors (including the incoming angle, the distance between the swimmer and the wall and the squirmer parameter $\beta$)} before escaping from the surface, while  squirmers with larger values of $|\beta|$ are weakly attracted to the surface following  periodic bouncing trajectories~\cite[]{li2014hydrodynamic, shen2018hydrodynamic}.   In comparison, a pusher squirmer in a viscoelastic fluid  can be permanently trapped   near the surface at high Deborah numbers due to the large   forces generated by  polymer extension behind the swimmer~\cite[]{li2014effect} (figure~\ref{fig:interaction}$a$), with pullers less affected. Similar results are also observed for finite-length undulatory swimmers~\cite[]{li2017near}.  \textcolor{black}{A theoretical analysis of a squirmer in a weakly viscoelastic fluid ($De\ll1$) shows that, depending on the initial orientation and distance of the swimmer from the wall, its near-wall residence time may increase or decrease, and the near-wall periodic trajectories turn into unstable spirals~\cite[]{yazdi2015swimming}. No permanent wall attraction of the swimmer was however observed in this analysis.}
For a squirmer actuating the fluid with a time-reversible small-amplitude slip velocity, a perturbation analysis predicts a permanent wall attraction for both pushers and pullers initially either swimming close to the wall~\cite[]{yazdi2014locomotion} or being oriented in a specific range of angles relative to the wall~\cite[]{yazdi2017effect}. 
   
Many of these results may differ when  non-hydrodynamic effects are included. For a self-propelled particle influenced by   Brownian noise, adding polymers into the fluid actually reduces its wall-trapping time~\cite[]{narinder2019active}  because the presence of polymers increases its rotational motion (figure~\ref{fig:interaction}$b$)~\cite[]{gomez2016dynamics, qi2020enhanced}.

\subsection{Interactions with external flows}

The    reciprocal theorem was used to compute the motion and hydrodynamic forces of a microswimmer in an arbitrary background flow in   a weakly non-Newtonian fluid~\cite[]{elfring2017force}. The final expression resembles the results in Eqs.~(\ref{eq:reciprocal1}) and (\ref{eq:reciprocal2}) with additional contributions from the background flow. A similar approach was applied to a spherical squirmer in a viscoelastic shear flow at low $De$ and showed that  it induced a secondary rotation of the swimmer with angular velocity of magnitude O($De\beta$)~\cite[]{de2017dynamics}.  
 At high $De$, further numerical simulations showed that a neutral squirmer always realigns its swimming direction with  the vorticity axis.  
 Pushers and pullers have more complex dynamics; pushers align with the vorticity axis if they have a small swimming speed and relatively low $De$, otherwise they reorient into the shear plane and tumble around the vorticity axis;   pullers display the opposite behavior~\cite[]{de2017dynamics}.  This feature might be exploited to separate   swimming microorganisms based on their propulsion mechanisms.

The   dynamics of microorganisms suspended in an external flow of viscoelastic fluids was also addressed using simplified point-swimmer models~\cite[]{ardekani2012emergence, mathijssen2016upstream}.  Assuming weak viscoelasticity, these studies describe the orientation of a discrete swimmer using    (Newtonian) Jeffery's orbits for an inert particle~\cite[]{jeffery1922motion} and incorporating a viscoelasticity-induced   migration~\cite[]{chan1997note} into the swimmer velocity. In a vortical flow,   viscoelasticity   leads to an aggregation of swimming microorganisms and dynamics along  limit cycles, whose sizes and shapes depend on both swimmer motility and fluid viscoelasticity~\cite[]{ardekani2012emergence} (figure~\ref{fig:interaction}$c$).  
In a Poiseuille flow, the oscillating upstream-swimming trajectory of  swimmers in  Newtonian fluids turns into damped oscillatory motion approaching the centerline of the channel in a weakly viscoelastic fluid with shear-dependent viscosity~\cite[]{mathijssen2016upstream}. \textcolor{black}{Other phenomena reported for spermatozoa in Newtonian fluids, such as cross-stream migration in  Poiseuille flows~\cite{ kumar2019effect} or buckling in an extensional flow~\cite{kumar2019flow}  may also be affected by viscoelasticity fluid.}

\subsection{Interactions between microswimmers}
Biological organisms and spermatozoa often swim together,  forming clusters and influencing one another through hydrodynamic interactions.  For example, adjacent spermatozoa are observed to swim cooperatively with their flagella beating in synchrony, leading to higher waving frequency and swimming speed~\cite[]{woolley2009study}.  
In a Newtonian fluid, a front-back asymmetry in the flagellar waveform is required for two infinitely-long flagella      to synchronise~\cite[]{elfring2009hydrodynamic}; the final synchronised state (in phase or opposite phase) is then just a function of the flagellar geometry.  
In contrast,  in a viscoelastic fluid pairwise interaction between two symmetric flagella  always leads to   in-phase synchronization, and is associated with the lowest energy dissipation in the surrounding fluid~\cite[]{elfring2010two}. 
Asymptotic analysis in the limit of a small dimensionless amplitude $A$ shows that   synchronization in a viscoelastic fluid occurs on a time scale $t\sim A^{-2}$, which is much faster than the typical phase-lock time in a Newtonian fluid $t\sim A^{-4}$~\cite[]{elfring2010two}.  Numerical simulations further showed that the speed and power efficiency for two synchronized flagella    are lower than in the case of a single flagellum, a conclusion valid for both Newtonian and viscoelastic fluids~\cite[]{chrispell2013actuated}.

In the case of multiple microswimmers,  viscoelasticity in the fluid can also impact  collective behavior. Bozorgi and Underhill  analyzed the linear stability of a homogeneous and isotropic distribution of microswimmers in a viscoelastic fluid using a continuum mean-field theory~\cite[]{bozorgi2011effect, bozorgi2013role}. \textcolor{black}{The average velocity of an infinitesimal volume of swimmer suspension is the sum of the swimming velocity for an isolated swimmer, the velocity in the fluid and a diffusive contribution for the center of mass of the swimmer cluster, while the average angular velocity is given by the sum of  Jeffery's reorientation and rotational diffusion.} The presence of a viscoelastic stress in the fluid is seen to have two influences: the extra polymer viscosity $\mu^p$ stabilizes the suspension, while the linear relaxation of the polymeric stress   destabilizes it. Similar to the Newtonian cases, a puller suspension  is always stable, while a pusher suspension is unstable. In a Newtonian fluid, the most unstable mode occurs at zero wavenumber and is of zero wavespeed. Increasing the value of $De$ increases its growth rate and generates a new peak of maximum growth rate at a non-zero wavenumber associated with non-zero wavespeed (see figure~\ref{fig:interaction}$d$). This result indicates that   viscoelasticity reduces the size of the coherent vortex structures induced by the collective motion of pusher swimmers, as was later confirmed in simulations~\cite[]{bozorgi2014effects, li2016collective}.

Numerical simulations of rod-like pusher swimmers showed that   viscoelasticity enhances   aggregation   by  generating  strong polymer stresses inside the gap between two swimmers parallel to each other (figure~\ref{fig:interaction}$e$)~\cite[]{li2016collective}. 
A  similar effect is observed for bovine sperm cells in viscoelastic fluids~\cite[]{tung2017fluid} (figure~\ref{fig:microswimmers}$c$). Pullers are less affected by fluid viscoelasticity, and instead the swimmers stick together at their ends like an asterisk.

For an active nematic suspension~\cite[]{hemingway2015active, hemingway2016viscoelastic}, adding polymers enhances the   stability of a system where polymer and nematic fields are coupled only by fluid velocity, mainly due to the increase of the viscosity. A direct coupling between polymers and nematics destabilizes the system and generates rich dynamics and flow states. At high $De$, viscoelasticity reduces drag in the transient active turbulence state by generating  strong polymer stresses acting against the local extensional flow and freezing the  pattern of defects in the flow. 

 More recently, cell division and motility in a viscoelastic fluid environment were considered theoretically using a two-phase fluid model and polymers were found to suppress the flow at the interface and damp the interfacial instabilities~\cite[]{emmanuel2020active}.

\section{Conclusion and Perspective}
\label{sec:conclusion}

In this review, we summarized our knowledge on microswimming in complex viscoelastic fluids, with an emphasis on mathematical theory, engineering applications and biological relevance. Complex  fluids, which  typically have  multi-scale structures, are involved in many biological processes so the  interactions between microswimmers and complex fluids might have played an important role in the evolution of organisms. Past research in this field, within the continuum frameworks of low-Reynolds number hydrodynamics and non-Newtonian rheology, provides fundamental physical insight in the complex behaviors of motile microorganisms in nature. By introducing multiple nonlinearities,  viscoelasticity in the fluid breaks the time-reversal symmetry of Newtonian Stokes flows, thereby changing the dynamics of microswimmers and their hydrodynamic interactions with their environment.

We summarized in detail the impact of viscoelasticity on the mobility of a single microswimmer and many of these results can be related to classical knowledge in the dynamics of suspensions in complex fluids.   For a spherical squirmer, the velocity change at small $De$ is  due to the asymmetric distribution of polymer stresses. At large $De$, strong polymer stresses in the local extensional regions reduce the speed, an effect that is reminiscent of the motion of passive particle and droplets in   viscoelastic fluids~\cite[]{chhabra2006bubbles}.  A small-amplitude flagellum undergoing  waving motion swims slower in a viscoelastic fluid because it creates a small-amplitude oscillatory shear flow and thus   experiences a reduced dynamic viscosity~\cite[]{bird1987dynamics, morrison2001understanding}.  For   models of flagella  with more realistic features (flexibility, finite-length, asymmetric waving form, etc.), the speed change is more intricate and depends on the details of both swimmer and fluid \cite{li2017near}.

The viscoelasticity of  the fluid also affects hydrodynamic interactions,  similarly to many other viscoelastic flows.  Viscoelasticity-induced surface/swimmer and swimmer/swimmer attractions for pushers resembles the surface attraction of a settling particle near a wall and the mutual attraction between settling particles in viscoelastic fluids~\cite[]{joseph1994aggregation, ardekani2007motion, ardekani2008two}. Aggregation and alignment occur for both swimmers and passive particles in a vortical viscoelastic flow~\cite[]{joseph1994aggregation, feng1996motion}. The role of polymer in destabilizing  swimmer suspensions is reminiscent  of  the occurrence of viscoelastic turbulence at negligible Reynolds numbers~\cite[]{groisman2000elastic, groisman2001efficient}. Future studies will be necessary to reveal the extent to which these phenomena are fundamentally identical and how   the swimmer motility affect these results. For example, numerical methods used to study suspensions of interacting particles and cells in viscoelastic fluids~\cite{li2015dynamics, raffiee2017elasto, raffiee2019numerical, raffiee2019suspension, man2017} could be extended to investigate the role of fluid elasticity on the transport of microswimmers.

\textcolor{black}{
As this review makes clear, there remain many important questions to   answer in future studies of microswimming in complex fluids. One key challenge will be to distinguish the universal effects of viscoelasticity on microswimming from effects that are specific to particular swimming modes. An interdisciplinary bridge between fluid mechanics, microbiology, and rheology will be necessary to build the full physical picture. One promising avenue is the extension to microswimmers of our  current understanding of hydrodynamics and rheology of  suspensions of passive particles in simple flows~\cite[]{leal1975slow, chan1977note, d2017particle}; we refer readers to a recent review of the rheology of suspensions in viscoelastic fluids~\cite[]{shaqfeh2019rheology}. Due to the nonlinearity in the case of complex fluids, the classical framework of Stokesian Dynamics  cannot be directly applied for viscoelastic fluids beyond the weakly nonlinear regime so numerical simulations able to fully resolve  interactions between microstructure  and microswimmer will therefore be needed. Another topic of potential great interest is the comparison between microswimming in polymer solutions and liquid crystals which have non-isotropic fluid elasticity due to the director field~\cite[]{stark2001physics, zhou2014living}. Inspired by active and passive microrheology~\cite[]{squires2005simple, vazquez2012sph}, we could also envision a situation where synthetic microswimmers become useful as microrheological probes for complex fluids. Once we fully unravel the dynamics of microswimmers in complex fluids, we can envision  practical applications where  synthetic microswimmers can be used in targeted drug delivery and in vivo biochemical detection studies.}

\section{Acknowledgments}
This project has received funding  from the National Science Foundation (Grants No.~CBET-1604423 and No.~CBET-1700961 to AMA) and the European Research Council (ERC) under the European Union's Horizon 2020 research and innovation programme  (grant agreement 682754 to EL). G. Li is supported by Scientific Research Staring Foundation 
(Grant No.~WH220401009)

\bibliographystyle{elsarticle-num}

\end{document}